\documentclass[12pt]{article}
\usepackage{epsfig,amssymb,euscript,bbold}
\usepackage{amsfonts}
\usepackage{eufrak}
\usepackage{bm}
\usepackage{amsmath}
\newcommand{\sss}{\sqrt{1-\lambda^3\rho^2}}
\def\be#1\ee{\begin{equation}#1\end{equation}}
\newcommand{\bea}{\begin{eqnarray}}
\newcommand{\eea}{\end{eqnarray}}
\newcommand{\ba}{\begin{array}}

\newcommand{\ea}{\end{array}}

\def\bbox{{\,\lower0.9pt\vbox{\hrule \hbox{\vrule height 0.2 cm
\hskip 0.2 cm \vrule height 0.2 cm}\hrule}\,}}
\newcommand{\dsl}{\pa \kern-0.5em /}

\newcommand{\G}{\Gamma}

\newcommand{\nn}{\nonumber \\}

\def\w{\wedge}
\def\mbb{\mathbb{R}}

\def\l{\lambda}
\def\s{\sigma}
\def\r{\rho}

\def\e{\epsilon}
\def\d{\delta}
\def\sss{\sqrt{1-\l^{3/2}\r^2}}

\def\ds{\raise.15ex\hbox{/}\kern-.57em\partial}
\def\Ds{\,\raise.15ex\hbox{/}\mkern-13.5mu D}

\def\D{\Delta}

\newcommand{\dd}{\mathrm{d}}
\newcommand{\ee}{\mathrm{e}}

\begin{document}

\makeatletter
\renewcommand{\theequation}{\thesection.\arabic{equation}}
\@addtoreset{equation}{section}
\makeatother

\baselineskip 18pt

\begin{titlepage}

\vfill

\begin{flushright}
Imperial/TP/2006/OC/02\\
\end{flushright}

\vfill

\begin{center}
   \baselineskip=16pt
   {\Large\bf Supersymmetric wrapped membranes,\\$AdS_2$ spaces, and
     bubbling geometries}
   \vskip 2cm
      Ois\'{\i}n A. P. Mac Conamhna and Eoin \'{O} Colg\'{a}in
   \vskip .6cm
      \begin{small}
      \textit{Theoretical Physics Group, Blackett Laboratory,\\
        Imperial College, London SW7 2AZ, U.K. \vskip .6cm The Institute for
        Mathematical Sciences,\\Imperial College, London SW7 2PE, U.K.}
        \end{small}
   \end{center}

\vfill

\begin{center}
\textbf{Abstract}
\end{center}

\begin{quote}
We perform a systematic study, in eleven dimensional supergravity, of
the geometry of wrapped brane configurations admitting $AdS_2$
limits. Membranes wrapping holomorphic curves in
Calabi-Yau manifolds are found to exhibit some novel features; in
particular, for fourfolds or threefolds, the gravitational effect of the
branes on the overall transverse space is only weakly restricted by the
kinematics of the Killing spinor equation. We also study the $AdS_2$
limits of the 
wrapped brane supergravity descriptions. From the description of
membranes wrapped in a two-fold, we derive a set of $AdS_2$ supersymmetry
conditions which upon analytic continuation coincide precisely with
those for the half-BPS bubbling geometries of LLM. From the
near-horizon limit of membranes wrapped in a three-fold, we obtain a
set of supersymmetry conditions which upon analytic continuation
describe a class of spacetimes which we identify as quarter-BPS
bubbling geometries in M-theory, with $SO(4)\times SO(3)\times U(1)$
isometry in Riemannian signature. We also study fivebranes wrapping a
special lagrangian five-cycle in a fivefold, in the presence of
membranes wrapping holomorphic curves, and employ the wrapped brane
supersymmetry conditions to derive a classification of the general
minimally supersymmetric $AdS_2$ geometry in M-theory.

\end{quote}

\vfill

\end{titlepage}

\setcounter{equation}{0}

\section{Introduction}
The AdS/CFT correspondence \cite{malda} has provided an unparalleled
theoretical laboratory in which ideas about quantum gravity, and their
dual field theory manifestations, may be explored in a controlled
mathematical framework. Though the best-understood examples of the
duality are for supersymmetric four-dimensional field theories, and
their $AdS_5$ duals in IIB string theory, the total space of AdS/CFT
duals in string and M-theory is vast, and remains relatively unexplored,
even for supersymmetric examples of the duality. This work is part of
an ongoing project which aims to shed light on the general geometrical
properties of supersymmetric $AdS$ spacetimes in M-theory, the brane
configurations that can give rise to them, and, by the
correspondence, the associated general properties of all CFTs
with M-theory duals.

The techniques we employ to mine the supergravity side of the correspondence 
involve, as a first step, extracting all the conditions
restricting the geometry of supersymmetric brane configurations, and
their associated $AdS$ spaces, that are
contained in the Killing spinor equation of eleven dimensional
supergravity, and repackaging them in a geometrically transparent
way. In doing so, one arrives at a set of necessary and sufficient
conditions for a spacetime to admit the desired number of Killing
spinors of the required form. A very useful way to package the
information is to use the G-structures defined by the Killing
spinors. This G-structure classification scheme was first formalised
in the context of supergravity in \cite{ns5}; it has since been
applied in many contexts, including the classification of minimally
supersymmetric solutions of lower \cite{5d}-\cite{12} and
eleven-dimensional \cite{j1}, \cite{j} supergravities. A refinement of
this technique, developed for the classification of spacetimes with
extended supersymmetry, has been given in \cite{8}, \cite{pap}, and
employed in eleven dimensions \cite{pap}-\cite{20}, and IIB
\cite{21}-\cite{23}.

In this paper, we will be concerned with the application of these
classification techniques to wrapped brane configurations admitting
$AdS_2$ limits. In keeping with the general philosophy of AdS/CFT, one
would expect that every $AdS$ spacetime in M-theory should arise
as the decoupling limit of some brane configuration. Supersymmetric
$AdS$ spacetimes of different dimensionalities, preserving different
ammounts of supersymmetry, may be obtained as the near-horizon limit
of branes wrapped on supersymmetric cycles, for which a whole zoo of
possibilities exist; for a review, see \cite{24}. The central
importance of wrapped brane configurations, and their $AdS$ limits, to the AdS/CFT
correspondence, means that these spacetimes have attracted a great deal
of attention, using a variety of different approaches. Early
investigations, following the work of Maldacena and Nu\~nez \cite{25},
\cite{26}, focussed on studying the near-horizon limit of wrapped brane
spacetimes in certain special cases, which were accesible via a
lower-dimensional gauged supergravity ansatz; see, for example,
\cite{27}-\cite{30}. Wrapped brane configurations have been classified
directly in eleven dimensional supergravity, under a variety of
assumptions, by numerous different authors
\cite{31}-\cite{37}. Seperately, various classes of supersymmetric
$AdS$ spacetimes in M-theory have been classified, by inserting an
appropriately general $AdS$ ansatz directly into the Killing spinor
equation of eleven dimensional supergravity; minimally
supersymmetric $AdS_3$ in \cite{35}, minimal purely magnetic $AdS_4$
in \cite{38} and minimal $AdS_5$ in \cite{39}. Some refinements of
these $AdS$ classifications have also appeared; for $AdS_4$ in
\cite{40} and $AdS_5$ with $\mathcal{N}=2$ supersymmetry in
\cite{41}. 

Recently, in \cite{wrap}, an explicit link between these different
avenues of investigation was supplied, first by using general
G-structure classification techniques to obtain the supergravity
description of wrapped brane spacetimes, and then by systematically employing a
procedure first used in \cite{39} to take the $AdS$ limits. Specifically, \cite{wrap} was
concerned with providing a supergravity description of M5 branes
wrapped on supersymmetric cycles in manifolds of $G_2$, $SU(3)$ or
$SU(2)$ holonomy, and then using this description to derive the
supersymmetry conditions for the $AdS$ limits of the wrapped brane
configurations. 

In this paper, we will apply the techniques of \cite{wrap} to wrapped brane
configurations in M-theory admitting $AdS_2$ limits. We
will study fivebranes wrapped on special lagrangian (SLAG)
five-cycles in Calabi-Yau fivefolds, in the presence of membranes wrapping
holomorphic curves. We will also study membranes wrapping holomorphic
curves (or, in alternatative teminology, K\"{a}hler two-cycles) in
Calabi-Yau $n$-folds, $n=2,...,5$. There is one remaining
possibility for a cycle on which all spacelike dimensions of an
M-brane can wrap: for fivebranes, the product of a SLAG-3 with
a K\"{a}hler two-cycle in an $SU(3)\times SU(2)$ manifold, but we do
not study this here. 

From the supergravity description of fivebranes and membranes wrapped
in a fivefold, we obtain a
classification of all minimally supersymmetric $AdS_2$ geometries
in M-theory. The supersymmetry conditions we obtain in this case are
rather complicated. Our results for membranes wrapped on holomorphic
curves in Calabi-Yau $n$-folds are more transparent. Before discussing
them, it is worth reviewing the findings
of \cite{wrap} for fivebranes. For the probe brane
analysis, one looks at the background with metric
\bea
\dd s^2=\dd s^2(\mathbb{R}^{1,p})+\dd s^2(\mathcal{M}_{10-p-q})+\dd
s^2(\mathbb{R}^q),
\eea
where $\mathcal{M}_{10-p-q}$ has $G_2$, $SU(3)$ or $SU(2)$ holonomy. One then introduces a
probe fivebrane, extended along the $\mathbb{R}^{1,p}$ directions and
wrapped on a cycle in $\mathcal{M}_{10-p-q}$. The kappa-symmetry
projections for the probe, in every case, imply that
although the supersymmetry is reduced by half, the structure group
defined by the Killing spinors is preserved. This has a very important
consequence in constraining the supergravity description of the system
when backreaction is turned on - in particular, it means that the
almost product structure of the spacetime, and the local flatness of the
overall transverse space, is
protected by 
supersymmetry up to warping. The input for the metric and Killing spinors in
the derivation of the supergravity description is that the Killing
spinors define the same algebraic structures as for the probe branes,
are simultaneous eigenspinors of five independent projection operators,
and that the metric contains a warped Minkowski factor of the appropriate
dimensionality to represent the unwrapped brane worldvolume. The
metric ansatz for the supergravity description is thus
\bea
\dd s^2=L^{-1}\dd s^2(\mathbb{R}^{1,p})+\dd s^2(\mathcal{M}_{10-d}).
\eea
Nothing is assumed at all about the form of the metric on
$\mathcal{M}_{10-d}$, beyond its independence (together with that of
$L$) of the Minkowski coordinates. But then it is found that
supersymmetry {\it implies} that the backreacted metric for the
supergravity description admits an almost product structure, in other
words, is of the form
\bea
\dd s^2=L^{-1}\dd s^2(\mathbb{R}^{1,p})+\dd s^2(\mathcal{M}_{10-p-q})+L^2\dd
s^2(\mathbb{R}^q),
\eea
where now $\mathcal{M}_{10-p-q}$ is deformed away from $G$-holonomy
but still admits a $G$-structure. A point we wish to emphasise
is that the gravitational effect of the fivebranes on the overall
transverse space is restricted, by supersymmetry, to inducing a
warping by $L^2$ (this warp factor will generically depend on the
coordinates of the overall transverse space). So
supersymmetry protects, up to warping, both the almost product
structure of the spacetime and the flatness of the overall transverse
space.    

We have found the behaviour of membranes, derived under exactly the
same assumptions, to be quite
different. Membranes wrapping a 
cycle in a fivefold is a special case which does not illustrate these
new features, since of course there are no overall transverse
directions (though it does have the unusual property that membranes
may be wrapped without breaking any supersymmetry). The new effects are
most pronounced for membranes wrapping a holomorphic curve in a
four-fold. The relevant background in this case is
\bea
\dd s^2=-dt^2+\dd s^2(\mathcal{M}_8)+\dd s^2(\mathbb{R}^2),
\eea
where $\mathcal{M}_8$ has $SU(4)$ holonomy. As we shall see below from
the probe brane analysis, the kappa-symmetry projection for a probe
membrane, extended along the timelike direction and wrapping a
holomorphic curve, breaks half the supersymmetry but now {\it
  increases} the structure group defined by the
Killing spinors to $SU(5)$. Then, in deriving the supergravity
description of the system, there is no symmetry principle which can
protect the almost product structure of the spacetime, so, in contrast
to what happens for fivebranes, the backreaction of the membranes on
the overall transverse space is largely unconstrained. The Killing
spinors preserved by a probe membrane wrapped in a four-fold define
exactly the same algebraic structures as those for a probe membrane
wrapped in a five-fold. Therefore there is no symmetry principle which
allows us to distinguish the supergravity descriptions of the two
systems, and no symmetry principle to confine the gravitational
effects of membranes wrapped in a fourfold to deforming
$\mathcal{M}_8$ away from $SU(4)$ holonomy and warping the overall transverse space. The
backreaction is not completely unconstrained of course, since the
spacetime must still admit an $SU(5)$ structure, satisfying certain
conditions.   

This effect is still present, though in reduced form, for membranes
wrapped in a three-fold. In this case, the special holonomy of the
background is $SU(3)$. The kappa-symmetry projection for a probe
breaks half the supersymmetry, but again increases the structure group
defined by the Killing spinors, this time to $SU(3)\times SU(2)$. In
this case, the almost product structure of the spacetime is
preserved. However, we shall see that it is consistent with the
Killing spinor equation for the gravitational effect of the
membranes on the overall transverse space to both induce a
warping, and to deform the overall transverse space away from being
flat to being of $SU(2)$ holonomy. We find that in the supergravity
description, the metric is given by
\bea
\dd s^2=-\Delta^2\dd t^2+\dd s^2(\mathcal{M}_{SU(3)})+\D^{-1}\dd
s^2(\mathcal{M}_{SU(2)}),
\eea
where $\mathcal{M}_{SU(2)}$ is of $SU(2)$ holonomy, is independent of
the coordinates of $\mathcal{M}_{SU(3)}$, and $\mathcal{M}_{SU(3)}$
admits an $SU(3)$ structure. Again this is in contrast to the
behaviour observed for fivebranes, though the effect is not so pronounced
as for four-folds. 

Finally, for membranes wrapped in a two-fold, we will see from the
probe brane analysis that the structure group of the backgound is
preserved in the presence of the probe. Then, in
the supergravity description, we will see that the almost product
structure of the spacetime is preserved, as is the local flatness of
the overall transverse space, up to a warping. So in this case, the
supersymmetry of the configuration protects the geometry of the
overall transverse space from the gravitational effects of the
membrane to the same degree as that observed for fivebranes; the
metric in the supergravity description is given by
\bea
\dd s^2=-\Delta^2\dd t^2+\dd s^2(\mathcal{M}_{SU(2)})+\D^{-1}\dd
s^2(\mbb^6),
\eea
where $\mathcal{M}_{SU(2)}$ admits an $SU(2)$ structure.

We have found that the remaining supersymmetry conditions for wrapped
membranes may be expressed in a very simple universal form. For
fivefolds and fourfolds, in the supergravity regime the systems are
indistinguishable, and the 
supersymmetry conditions are the same; in terms of the holomorphic
five-form $\Omega$ and the almost complex structure $J$ which specify
the $SU(5)$ structure, they may be expressed as
\bea
\dd (e^0\w\mbox{Re}\Omega)&=&0,\nn
\dd\star J&=&0,\nn
F&=&-\dd(e^0\w J).
\eea
In the first of these equations we clearly recognise the generalised
calibration condition for a probe M5 brane on a SLAG cycle in the backreacted
geometry. In fact, this condition also implies $\dd
(e^0\w\mbox{Im}\Omega)=0$, and of course a SLAG cycle may be
calibrated by either the real or imaginary part of the holomorphic
form. We also observe that the generalised calibration for the
membrane worldvolume is required, by the four-form field equation, to
be a harmonic form in spacetime. For threefolds or twofolds, if we let
$\Delta^{n/2}I_{n}$, $n=2,3$, denote an arbitrary closed $n$-form on
the overall transverse space, the supersymmetry conditions we find may
be expressed as
\bea
\dd (e^0\w\Omega_{SU(n)}\w I_n)&=&0,\nn
\dd\star J_{SU(n)}&=&0,\nn
F&=&-\dd(e^0\w J_{SU(n)}).
\eea
Once again, the first of these equations is manifestly a generalised
calibration condition, for all the ways in which a probe fivebrane can
wrap the backreacted geometry while preserving supersymmetry. And again the
four-form field equation implies that the generalised calibration for
the membrane worldvolume is harmonic in spacetime.    

Once we have derived the wrapped brane supersymmetry conditions, we
study their $AdS_2$ limits. In taking the $AdS$ limit of a metric
which is the warped product of a timelike line with a Riemannian
ten-manifold, we must pick out the $AdS$ radial direction from the
ten-manifold. How we do this will be discussed in detail. For branes
wrapped in a fivefold, the limiting procedure is completely
general. Therefore our $AdS_2$ limit of the SLAG-5 supersymmetry
conditions gives a classification of the general minimally
supersymmetric $AdS_2$ spacetime in M-theory; from the $AdS_2$ limit
of membranes on a holomorphic curve, we obtain a classification of the
general minimally supersymmetric $AdS_2$ spacetime in M-theory with
purely electric fluxes. Our supersymmetry conditions in this case
coincide with those of \cite{nakwoo}, though we derive them without
assuming that the nine-manifold transverse to the $AdS$ factor is
compact. 

For membranes wrapped in fourfolds, there is nothing new to discuss in
taking the $AdS$ limit, since the wrapped brane supersymmetry
conditions are identical to those for fivefolds. But for threefolds
and twofolds, what we have found confirms and extends
the beautiful relationship between bubbling geometries and $AdS$ spaces proposed
by LLM \cite{41}, and also makes manifest the intimate link between
these geometries and the $SU(n)$ structures of the brane configurations. For membranes wrapped in a
twofold, our limiting procedure produces a set of supersymmetry
conditions for half-BPS $AdS$ spacetimes. In addition to the $AdS$
isometries, supersymmetry implies that these spacetimes will have
$U(1)\times SO(6)$ isometry. The $SO(6)$ comes from the sphere in the
overall transverse space, whose isometries are promoted to isometries
of the full solution in our limit. The $U(1)$ arises because the $AdS$
radial direction comes partly from the overall transverse space and
partly from $\mathcal{M}_{SU(2)}$ in the wrapped brane metric; the
part that lies in $\mathcal{M}_{SU(2)}$ is paired with a vector
$\hat{w}$, which generates the $U(1)$, by $J_{SU(2)}$. Analytically
continuing the metric and supersymmetry conditions to $AdS_5$, or to the
bubbling geometries where the $U(1)$ is timelike, our conditions map
exactly to those of LLM. The $AdS_5$ conditions of LLM were derived in
\cite{wrap} by taking the $AdS$ limit of the supergravity description
of fivebranes wrapped on holomorphic curves in twofolds.  

For membranes wrapped in threefolds, we study the $AdS$ limit when the
overall transverse $SU(2)$ manifold is chosen to be $\mbb^4$. We find
there are two ways of taking the $AdS$ limit. One is when the $AdS$
radial direction is assumed to lie entirely in the overall transverse
space. In this case, the limit is $AdS_2\times CY_3\times S^3$. The
$AdS_3$ limit of fivebranes wrapped on K\"{a}hler four-cycles in
threefolds, again when the $AdS$ radial direction is assumed to lie
entirely in the overall transverse space of the wrapped brane metric,
is $AdS_3\times CY_3\times S^2$. So again the supersymmetry
conditions for the $AdS$ limit of the membrane configuration
analytically continue to those of the fivebrane configuration. But in
this case there is no additional $U(1)$ isometry, so a bubbling interpretation is unclear.  
However, there is another $U(1)$ in the second way of taking the $AdS$
limit of the wrapped membranes, where the $AdS$ radial direction is
assumed to lie partly in the overall transverse space and partly in
the $SU(3)$ structure manifold of the wrapped brane metric. It has
exactly the same origin as in the half-BPS case: it is the vector
paired with the part of the $AdS$ radial direction lying in
$\mathcal{M}_{SU(3)}$ by $J_{SU(3)}$. Similarly, in \cite{wrap} it was
found that there is an extra $U(1)$ isometry in the $AdS$ limit of
fivebranes on four-cycles in threefolds, when the $AdS$ radial
direction comes partly from the threefold and partly from the overall
transverse space. And we have found that the conditions we derive here
for the $AdS$ limit of the membranes analytically continue precisely
to those of \cite{wrap} for the $AdS$ limit of the fivebranes. But now
these spacetimes admit another analytic continuation; in Riemannian
signature the isometry is $U(1)\times SO(3)\times SO(4)$, and
continuing so that the $U(1)$ is timelike, we get spacetimes that we
interpret as the quarter-BPS bubbling geometries of M-theory. We are
unaware of any explicit known solutions of the supersymmetry
conditions in this case, but it will be very interesting to study the
supersymmetry conditions in more detail.

At this point it is worth briefly reviewing the literature on the brane
configurations we study, and also supersymmetric $AdS_2$ spacetimes in M-theory. A gauged supergravity
investigation of the near-horizon limits in some special cases cases
was given for fivebranes on SLAG five-cycles in \cite{29}, and for
membranes on holomorphic curves in \cite{42}. The supersymmetry
conditions for a single timelike Killing spinor in eleven dimensions
(as appropriate for the description of M5 branes on SLAG five-cycles)
were first given in \cite{j1}. Membranes wrapping holomorphic curves
in Calabi-Yau manifolds were studied, using the Fayyazuddin-Smith
ansatz and from the perspective of generalised calibrations, in
\cite{36} and \cite{43}. The conditions for minimally supersymmetric
$AdS_2$ spacetimes in M-theory, with vanishing magnetic flux and
compact internal space, were
derived in \cite{nakwoo}. And of course the supersymmetry conditions for a
class of half-BPS $AdS_2$ spacetimes in M-theory may be derived by
analytic continuation of the results of LLM \cite{41}.

In this paper, we have not attempted to find any explicit new
solutions of the supersymmetry conditions, and of course, performing
classifications along the lines of those given here is only the first
step in exploring the space of AdS/CFT duals 
in M-theory. To provide new explicit concrete examples of the duality,
or new explicit bubbling solutions,
the supersymmetry conditions (and the Bianchi identity/ field
equations, where appropriate) must be solved on the supergravity side. However, the geometrical
insight provided by the G-structure formalism has proven extremely useful
in integrating the supersymmetry conditions. The by-now celebrated $Y^{p,q}$
spaces were constructed directly from the results of the $AdS_5$
classification of \cite{39}; the field theory duals have been
identified \cite{44} and much further progress has been made. Some
properties of the duals of other explicit $AdS_5$ solutions found in
\cite{39} have also very recently been elucidated
\cite{45}. Furthermore, inspired by the insight provided by the
results of G-structure classifications, an extremely rich
five-parameter family of supersymmetric $AdS_3$ solutions of M-theory
has recently been constructed in \cite{46}. These solutions are dual
to field theories with $\mathcal{N}=(2,0)$ supersymmetry, and arise as
the near-horizon limits of M5 branes wrapping K\"{a}hler four-cycles
in Calabi-Yau four-folds. Eight doubly-countably infinite compact
families of these solutions, which may be dimensionally reduced and
T-dualised to IIB, were studied in \cite{47}, and the central charges of
the CFT duals computed. It is to be hoped that completing the
classification of wrapped brane configurations and their $AdS$ limits
in M-theory may facilitate similar progress in the future.

The plan of the rest of this paper is as follows. In section 2, we
study branes wrapping supersymmetric cycles in Calabi-Yau
five-folds. From the supergravity description of fivebranes wrapping
SLAG five-cycles together with membranes wrapping K\"{a}hler
two-cycles, we derive the supersymmetry conditions for a general
minimally supersymmetric $AdS_2$ spacetime in M-theory. We also show
how the supersymmetry conditions for a minimally supersymmetric
$AdS_2$ spacetime with purely electric fluxes \cite{nakwoo} may be
obtained directly from the supersymmetry conditions for membranes
wrapped on a K\"{a}hler two-cycle.

In section 3, we study membranes wrapping K\"{a}hler two-cycles in
Calabi-Yau $n$-folds, $n=2,3,4$, performing a probe brane analysis and
then deriving the supergravity description of the wrapped brane
configurations. 

In section 4, we study the $AdS$ limits of these configurations. We
describe our limiting procedure in detail, and employ it to derive the
$AdS$ supersymmetry conditions from those of the wrapped branes.

Section 5 concludes. Miscellaneous technical material is relegated
from the main body of the text to several appendices. Throughout the
text we use all the spinorial conventions of \cite{j1}.

\section{Branes wrapped on cycles in $SU(5)$ manifolds}\label{oop}
In this section, we will study the supergravity description of branes
wrapped on cycles in Calabi-Yau five-folds, together with their
near-horizon limits. We are interested in two configurations:
fivebranes wrapped on SLAG five-cycles in the presence of membranes on
holomorphic curves; and secondly, configurations just with wrapped
membranes. We will perform a brief probe-brane
analysis, to identify the Killing spinors preserved in each case. Then
we will discuss the supersymmetry conditions for the wrapped branes,
from which we will finally derive the supersymmetry conditions for the
$AdS$ limits. Throughout this paper, all Killing spinors are
timelike. We will therefore use a timelike spacetime
basis, given by
\bea
ds^{2} = -(e^{0})^{2} + \d_{ab} e^{a} e^{b},
\eea
where $a,b = 1,..,9,\sharp$, in the following.

\subsection{Probe branes}
We are interested in probe branes on supersymmetric cycles in
$\mbb\times CY_5$. We may choose the Killing spinors preserved by this
background to be the pair of $SU(5)$ invariant spinors satisfying
\bea\label{proj2}
\G^{1234} \eta = \G^{3456} \eta = \G^{5678} \eta = \G^{789 \sharp} \eta
= - \eta.
\eea
From the spinor bi-linears, we may construct the complex structure $J$
and also the holomorphic five-form $\Omega$, given by
\bea
J &=& e^{12} + e^{34} + e^{56} + e^{78} + e^{9\sharp},\nn 
\Omega&=&(e^1+ie^2)(e^3+ie^4)(e^5+ie^6)(e^7+ie^8)(e^9+ie^\sharp).
\eea
Observe that the special
holonomy projections imply that we can wrap a membrane for free on a
K\"{a}hler two-cycle, calibrated by $J$, in the ten-manifold. We can
take the membrane kappa-symmetry projection to be  
\bea
\G^{012}\eta=-\eta.
\eea
Therefore the two Killing spinors preserved by a membrane wrapping a
holomorphic curve define an $SU(5)$
structure with two supersymmetries.
Now consider introducing probe M5 branes wrapped on supersymmetric SLAG
five-cycles. Such cycles are, by definition, calibrated 
by $\mbox{Re}\Omega$. For such a cycle, we choose the projection 
\bea\label{m2proj}
\G^{013579} \eta = -\eta.
\eea
This projects out one of the Killing spinors of the $CY_{5}$, so these
backgrounds preserve a single supersymmetry. We will reserve the
notation $\xi$ for a timelike spinor satisfying the projections
(\ref{proj2}) and (\ref{m2proj}). The second $SU(5)$ Killing spinor, projected
out by (\ref{m2proj}), is
\bea
\frac{1}{10}J_{ab}\G^{ab}\xi=\G^0\xi.
\eea
Taking $\xi$ to have unit norm, the forms it defines in eleven dimensions are
\bea
K&=&\overline{\xi}\G^{(1)}\xi=-e^0,\nn
\Theta&=&\overline{\xi}\G^{(2)}\xi=J,\nn
\Sigma&=&\overline{\xi}\G^{(5)}\xi=\frac{1}{2}e^0\wedge J\w
J+\mbox{Re}\Omega.
\eea

\subsection{The supergravity description}
Demanding the existence of Killing spinors defining the same algebraic
structures as for the probe branes, and which are simultaneous
eigenspinors of the projection operators of \eqref{proj2},
\eqref{m2proj}, it is now an easy matter to give
the supersymmetry conditions in the supergravity description. In going
away from the probe brane approximation, the backreaction of the
branes will deform the Calabi-Yau away from $SU(5)$ holonomy, but
will still preserve $SU(5)$ structure. Let us now briefly state our bosonic
ansatz. We demand that our wrapped brane metric is of 
the form 
\be
\label{timelikeansatz}
\dd s^{2} = -\Delta^{2} \dd t^{2} + h_{MN}\dd x^{M}\dd x^{N}.
\ee
Our timelike frame is
realised by  $e^{0} = \Delta \dd t$ and $e^{a} = e^{a}_{M}\dd x^{M}$,  where
we refer to the ten-manifold spanned by $e^{a}$ as the base $B$. Since
we are interested in wrapped brane configurations admitting $AdS$
limits, we have required that the timelike direction is not fibred
over the base. The $t$-independence of $\Delta$ and $h$ follows from
the supersymmetry conditions of \cite{j1}. Our ansatz for the flux is 
\be F = \Delta^{-1} e^{0} \w H + G, 
\ee
where $H$ is a three-form and $G$ is a four-form defined on $B$. We
demand that $H$ and $G$ are independent of $t$. Now we give the
supersymmetry conditions.

\subsubsection{Fivebranes on SLAG five-cycles}
In this case with a single timelike
spinor, the supersymmetry conditions may be obtained simply by truncating the results of \cite{j1}
to our ansatz. Turning on backreaction
induces a warping of the timelike direction, and the forms defined by
the Killing spinor will rescale
\bea
K&=&-\D e^0,\nn
\Theta&=&\D J,\nn\label{resca}
\Sigma&=&\D\Big(\frac{1}{2}e^0\w J\w J+\mbox{Re}\Omega\Big).
\eea
The only restriction on the base $SU(5)$ structure implied by
supersymmetry is 
\be
\label{restrict}
\mbox{Re}\Omega \lrcorner \dd\mbox{Re}\Omega = 8 \dd \log \Delta . 
\ee
The four-form field strength is then given by
\bea
F&=&-\dd(e^0\w
J)+\frac{1}{2}\star\dd(e^0\w\mbox{Re}\Omega)-\frac{1}{2}W_1\w
J+\frac{1}{4}(\dd\log\D+W_4)\lrcorner\mbox{Im}\Omega\nn\label{SU5fourform}&&+F^{\mathbf{75}}\nn
&=&-\dd(e^0\w J)
-\frac{1}{2}W_2-\frac{1}{3}W_1\w J+\frac{1}{4}(\dd
\log\D+W_4)\lrcorner\mbox{Im}\Omega+F^{\mathbf{75}},
\eea
where $F^{\mathbf{75}}$ is a four-form defined on the base in the
$\mathbf{75}$ of $SU(5)$ which is unfixed by the supersymmetry
conditions. The $SU(5)$ torsion modules $\mathcal{W}_i$, $i=1,...,5$, are
defined by
\bea
\dd J&=&\frac{1}{8}\mathcal{W}_1\lrcorner\mbox{Im}\Omega+\mathcal{W}_3+\frac{1}{4}\mathcal{W}_4\w J,\nn\label{213}
\dd\mbox{Re}\Omega&=&\frac{1}{6}\mathcal{W}_1\w J^2+\mathcal{W}_2\w
J+\frac{1}{8}\mbox{Re}\Omega\w \mathcal{W}_5.
\eea
At this point, we should clarify a potentially misleading aspect of
our terminology. We refer to the supersymmetry conditions given above as
``wrapped brane supersymmetry conditions''; however, they are
sufficiently general that in addition to wrapped brane configurations,
they are solved by many supersymmetric
spacetimes which do not contain any branes at all. A trivial example
is flat space with vanishing flux. On the other hand, the wrapped
brane supersymmetry conditions will indeed be
solved by all wrapped brane configurations in M-theory which admit an
$AdS_2$ limit. With this understanding - that what we refer to as
wrapped brane supersymmetry conditions are solved by all wrapped brane
configurations in the desired class, but also admit other solutions -
we will continue to use this terminology throughout the paper. And of
course, none of the solutions to the wrapped brane supersymmetry
conditions which do not in fact describe brane configurations will
admit an $AdS$ limit. 

\subsubsection{Membranes on holomorphic curves}\label{kk}
The supersymmetry conditions for membranes wrapped on holomorphic
curves may be derived simply by setting the magnetic flux to zero in
(\ref{SU5fourform}). With the metric (\ref{timelikeansatz}), the
resulting equations may be succinctly expressed in the form given in
the introduction,
\bea
\dd (e^0\w\mbox{Re}\Omega)&=&0,\nn
\dd\star J&=&0,\nn\label{electricsu5}
F&=&-\dd(e^0\w J).
\eea
To verify that these configurations indeed admit two supersymmetries,
observe that given an $SU(5)$ structure defined by \eqref{resca}
satisfying \eqref{electricsu5}, we may always define a second $SU(5)$
structure, satisfying \eqref{electricsu5}, according to
\bea
K'&=&\D e^0,\nn
\Theta'&=&-\D J,\nn
\Sigma'&=&\D\Big(-\frac{1}{2}e^0\w J\w J+\mbox{Re}\Omega\Big).
\eea
These are the forms associated to the spinor $\Delta^{1/2}\G^0\xi$,
which is thus Killing.

\subsection{The $AdS$ limits}\label{rrr}
In this subsection, we will take the general $AdS$ limits of the
wrapped brane metrics and supersymmetry conditions of the previous
subsections. To do so, we observe that a warped $AdS_2$ product
metric may be viewed as a special case of the metric (\ref{timelikeansatz}), if we
write the $AdS$ metric in Poincar\'{e} co-ordinates:
\be 
\frac{1}{m^{2}} \dd s^{2}(AdS_{2}) = -e^{-2mr}\dd t^2 +
\dd r^{2}. 
\ee 
Therefore to make contact with \eqref{timelikeansatz}, we require that 
\be
\Delta = e^{-mr}\l^{-1/2}.
\ee 
We demand that the $AdS$ warp factor $\l$, and the frame on the space
transverse to the $AdS$ factor, 
are independent of the $AdS$ coordinates. To complete the $AdS$ metric
ansatz, we must pick out the $AdS$ radial 
direction from the base space. Using the transitive action of $SU(5)$ on the base, we may choose
\bea
\l^{-1/2}\dd r=e^{\sharp}.
\eea
Picking out a preferred direction on the base associated to the
doubling of supersymmetry reduces the structure
group defined by the Killing spinors to $SU(4)$; the metric becomes
\bea
\dd s^2=\frac{1}{\l m^2}\dd s^2(AdS_2)+\dd
s^2(\mathcal{M}_8)+e^9\otimes e^9.
\eea
The $SU(4)$ structure is defined on $\mathcal{M}_8$. In terms of the
$SU(4)$ structure forms,
\bea
J_{SU(4)}&=&e^{12}+e^{34}+e^{56}+e^{78},\nn
\Omega_{SU(4)}&=&(e^1+ie^2)(e^3+ie^4)(e^5+ie^6)(e^7+ie^8),
\eea
the $SU(5)$ structure decomposes according to
\bea
J&=&J_{SU(4)}+\l^{-1/2}e^9\w \dd r,\nn
\mbox{Re}\Omega&=&\mbox{Re}\Omega_{SU(4)}\w
e^9-\l^{-1/2}\mbox{Im}\Omega_{SU(4)}\w \dd r,\nn
\label{41}\mbox{Im}\Omega&=&\mbox{Im}\Omega_{SU(4)}\w
e^9+\l^{-1/2}\mbox{Re}\Omega_{SU(4)}\w \dd r.
\eea

To complete the
$AdS$ limit, we demand that the only non-vanishing electric flux
contains a factor 
proportional to the $AdS$ volume form, and that the magnetic flux has
no components along the $AdS$ radial direction.

\subsubsection{The $AdS$ limit of fivebranes on SLAG five-cycles}
In this subsection, we will describe the result of this limiting
procedure as applied to the SLAG five-cycle; a more detailed
discussion of their derivation is given in appendix A. In
the $AdS$ limit there is an $SU(4)$ structure in nine dimensions; the forms
defining this structure are $e^9$, $J_{SU(4)}$ and
$\mbox{Im}\Omega_{SU(4)}$ (we could instead have chosen
$\mbox{Re}\Omega_{SU(4)}$, their exterior derivatives contain the same
information). For the remainder of this section, we will drop the
$SU(4)$ subscripts; it is understood that all forms and torsion
modules are now of $SU(4)$. The conditions on the intrinsic torsion,
in terms of the 
$SU(4)$ torsion modules in nine dimensions, may be expressed
as
\bea
\dd(\l^{-1/2}J)&=&0,\nn
\dd\mbox{Im}\Omega&=&\mathcal{W}_2^{(\mathbf{20}+\bar{\mathbf{20}})}\w
J+\frac{1}{4}\mbox{Im}\Omega\w\mathcal{W}_5^{(\mathbf{4}+\bar{\mathbf{4}})}\nn&+&\Big[\l^{1/2}m\mbox{Re}\Omega+\mbox{Im}\Omega\partial_9\log\lambda+\mathcal{W}_6^{(\mathbf{10}+\bar{\mathbf{10}})}\Big]\w
e^9,\nn\label{phto}
\dd
e^9&=&\mathcal{W}_7^{(\mathbf{6}+\bar{\mathbf{6}})}+\mathcal{W}_8^{\mathbf{15}}+\frac{1}{4}\mathcal{W}_9^{\mathbf{1}}J+\frac{1}{2}(3\tilde{\dd}\log\l-\mathcal{W}_5^{(\mathbf{4}+\bar{\mathbf{4}})})\w
e^9.
\eea
The numbering of the torsion modules is chosen to emphasise that many
modules are zero, and of those non-zero, many are not generic. We use
$\tilde{\dd}$ to denote the exterior derivative restricted to
$\mathcal{M}_8$. The flux is given in terms of the torsion modules by
\bea
F&=&\frac{1}{m^2}\mbox{Vol}_{AdS_2}\w[\dd(\l^{-1}e^9)-m\l^{-1/2}J]-\frac{1}{2}[\mathcal{W}_6+(\mathcal{W}_7\lrcorner\mbox{Re}\Omega)\w
J]\nn
&+&\frac{1}{4}(\mathcal{W}_9-m\l^{1/2})\mbox{Re}\Omega+\frac{3}{8}\mbox{Im}\Omega\partial_9\log\l+F^{\mathbf{20}}\nn\label{phtoo}
&-&\Big[J\cdot\mathcal{W}_2+\frac{1}{4}(\mathcal{W}_5-4\tilde{\dd}\log\l)\lrcorner\mbox{Im}\Omega\Big]\w
e^9,
\eea
where for an $n$-from $\Lambda$, we have defined
$J\cdot\Lambda_{i_1...i_n}=nJ_{[i_1}^{\;\;j}\Lambda_{|j|i_2...i_n]}$,
and $F^{\mathbf{20}}$ is a primitive $(2,2)$ form on $\mathcal{M}_8$
which is unfixed by the supersymmetry conditions. 

\subsubsection{The $AdS$ limit of membranes on holomorphic curves}
Now we will state the supersymmetry conditions we have derived for the
$AdS$ limit of membranes wrapping holomorphic curves; more details of
their derivation are given in appendix A. There are two equivalent ways in which
these supersymmetry conditions may be arrived at; first, by taking the $AdS$ limit of
\eqref{electricsu5} directly, and second, by setting the magnetic flux to
zero in the $AdS$ limit of the SLAG-5 supersymmetry conditions,
\eqref{phto}, \eqref{phtoo}. It serves as a useful consistency check
to verify that in commuting the order of these limits one indeed arrives at
the same answer, and we have done so. The metric is given by
\bea
ds^2=\frac{1}{\l
  m^2}[ds^2(AdS_2)+\l^{3/2}ds^2(\mathcal{M}_8)+(dz+\sigma)^2],
\eea
where $\partial_z$ is Killing and $\mathcal{M}_8$ is K\"{a}hler, with
Ricci form $\mathcal{R}$ and scalar curvature $R$ given by
\bea
\mathcal{R}&=&\dd\s,\\
R&=&2\l^{3/2}.
\eea
The flux is given by
\bea
\label{eflux}
F=\mbox{Vol}_{AdS_2}\w[\dd(\l^{-1}e^9)-m\l^{-1/2}J_{SU(4)}].
\eea
These conditions are identical to those of \cite{nakwoo}, though 
they are also valid for non-compact $\mathcal{M}_8$.

\section{Membranes wrapped on cycles in $SU(n)$ manifolds}
In this section, we will study the supergravity description of
membranes wrapping holomorphic curves in Calabi-Yau four, three and
two-folds. As discussed in the introduction, what we will see is that
supersymmetric wrapped membranes 
can affect the geometry of spacetime in a way that is qualitatively
different to that of wrapped fivebranes. 

As in the previous section, we will first perform a probe brane
analysis of the wrapped membrane configurations to determine the
preserved supersymmetries, and the $G$-structures associated to them,
and then derive the supergravity description of the same
configurations with the same supersymmetries. In performing the probe
brane and subsequent supergravity analysis, it is very useful to construct an
explicit spinorial basis which respects $SU(5)$ covariance. This is
one of the essential points of the refined $G$-structure formalism of
\cite{8}, \cite{pap} - decomposing the space of spinors into modules
of the structure group, and in so doing exploiting the geometrical
structure present to organise and render tractable extremely
complicated supergravity calculations. Using the timelike spacetime basis and
the fiducial timelike spinor $\xi$ defined in section \ref{oop}, we choose our
spinorial basis to be 
\bea
\xi,\;\;\G^0\xi,\;\;\G^a\xi,\;\;\frac{1}{4}A^{(\mathbf{10}+\bar{\mathbf{10}})}_{ab}\G^{ab}\xi.
\eea
Here $a,b=1,...,10$, and the $A^{(\mathbf{10}+\bar{\mathbf{10}})}$
furnish a basis for $(2,0)+(0,2)$ forms of $SU(5)$; explicitly, we may
choose the $A^{(\mathbf{10}+\bar{\mathbf{10}})}$ to be $e^{13}-e^{24}$,
$e^{14}+e^{23}$, etc. In choosing this basis, we are exploiting the
isomorphism between the space of Majorana spinors in eleven dimensions
and forms of $SU(5)$:
\bea
\mathbf{32}=(\mathbf{1}+\bar{\mathbf{1}})\oplus(\mathbf{5}+\bar{\mathbf{5}})\oplus(\mathbf{10}+\bar{\mathbf{10}}).
\eea
Observe that each member of this basis may be distinguished by its
eigenvalues under the projection operators of \eqref{proj2} and
\eqref{m2proj}. We will use this basis extensively in the following.

\subsection{Probe brane analysis}

\subsubsection{Probe membranes on holomorphic curves in fourfolds}
First we look at probe membranes wrapping a holomorphic curve in a
fourfold. The background is $\mathbb{R}^{1,2}\times CY_4$, where we take
the $\mathbb{R}^{1,2}$ to be spanned by $e^0$, $e^9$, and $e^{\sharp}$. The
probe membrane is extended along the timelike direction $e^0$, and
$e^9,\;\;e^{\sharp}$ span the overall transverse space. In the
absence of the probe membrane, the background preserves four
Killing spinors, which we may take to satisfy the projections 
\bea\label{proj3a}
\G^{1234} \eta = \G^{3456} \eta = \G^{5678} \eta 
= - \eta.
\eea
We may choose the four linearly independent solutions of these
projection conditions to be
\be
\begin{array}{cccc}
\xi, & \G^{0} \xi, & \G^{9} \xi, & \G^{\sharp} \xi.
\end{array}
\ee
These Killing spinors define an $SU(4)$ structure in eleven dimensions.
The introduction of a membrane probe into this background breaks half
of these supersymmetries. We must impose the kappa-symmetry projection
for the probe brane, which we may choose to be
\be\label{su:m2proj}
\G^{012} \eta = -\eta.
\ee
This projects out the Killing spinors $\G^{9}\xi,\;\; \G^{\sharp}\xi$; the
Killing spinors preserved by the background in the presence of the
probe brane are
\bea
\xi,\;\;\G^0\xi.
\eea
These Killing spinors define an $SU(5)$ structure; equivalently, they
are annihilated by an $SU(5)$ subalgebra of the Lie algebra of
$Spin(1,10)$.  Thus we arrive at the
surprising conclusion that wrapping a probe membrane in a Calabi-Yau
fourfold, in breaking half the supersymmetries, increases the
structure group defined by the Killing spinors from $SU(4)$ to
$SU(5)$.

\subsubsection{Probe membranes on holomorphic curves in threefolds} 
Now we look at probe membranes wrapped in a threefold. The background
is $\mathbb{R}^{1,4}\times CY_3$, where now the overall transverse
space is spanned by $e^7,...,e^{\sharp}$. In the absence of the probe
brane, the Killing spinors preserved by the background satisfy the
projections
\bea
\G^{1234}\eta=\G^{3456}\eta=-\eta,
\eea
and we may choose them to be
\bea
\xi,\;\;\G^0\xi,\;\;\G^7\xi,\;\;\G^8\xi,\;\;\G^9\xi,\;\;\G^{\sharp}\xi,\;\;\frac{1}{4}A_{ab}^1\G^{ab}\xi,\;\;\frac{1}{4}A_{ab}^2\G^{ab}\xi,
\eea
where
\bea\label{A}
A^1=e^{79}-e^{8\sharp},&&A^2=e^{7\sharp}+e^{89}.
\eea
These eight Killing spinors define an $SU(3)$ structure. Now,
introducing the membrane probe, we may again choose the kappa-symmetry
projection to be (\ref{su:m2proj}); this projects out
the four $\G^{a} \xi$, $a=7,...,\sharp$, Killing
spinors above. The surviving Killing spinors define an $SU(3) \times
SU(2)$ structure in eleven dimensions; the most general element of the
Lie algebra of $Spin(1,10)$ which annihilates all four is
\bea
B^{\mathbf{8}}_{ab}\G^{ab}+C^{\mathbf{3}}_{ab}\G^{ab},
\eea
where $B^{\mathbf{8}}$ is an arbitrary primitive (1,1) form (ie, in the
adjoint) of an $SU(3)$ acting on the 123456 directions, and
$C^{\mathbf{3}}$ is an arbitrary primitive (1,1) 
form of an $SU(2)$ acting on the $789\sharp$. Once again we observe
the feature that in  breaking half the supersymmetry, a probe membrane
increases the structure group of the background.

\subsubsection{Probe membranes on holomorphic curves in twofolds}  
Finally, we look at membranes wrapped in a twofold. In this case, the
background is $\mathbb{R}^{1,6}\times CY_2$, and we take the overall
transverse space to be spanned by $e^5,...,e^{\sharp}$. The sixteen Killing
spinors of the background satisfy the single projection
\bea
\G^{1234}\eta=-\eta,
\eea
and the sixteen basis spinors satisfying this projection may be easily
found. Introducing the probe brane, we must once again impose the
kappa-symmetry projection, which we again choose to be
\eqref{su:m2proj}. The eight surviving Killing spinors are
\bea
\xi,\;\;\G^0\xi,\;\;A^A_{ab}\G^{ab}\xi,\;\; A=1,...,6,
\eea
where $A^{1,2}$ are as defined in \eqref{A}, and
\bea
A^3=e^{59}-e^{6\sharp},&&A^4=e^{5\sharp}+e^{69},\nn
A^5=e^{57}-e^{68},&&A^6=e^{58}+e^{67}.
\eea
The most general element of $Spin(1,10)$ annihilating these eight
Killing spinors is
\bea
D_{ab}^{\mathbf{3}}\G^{ab},
\eea
where $D^{\mathbf{3}}$ is a primitive (1,1) form of an $SU(2)$ acting
on the 1234 directions. Therefore these Killing spinors define an
$SU(2)$ structure in eleven dimensions. In this case, the structure
group of the background is preserved under the introduction of the
probe brane.

\subsection{The supergravity description}
Now we will use the Killing spinors obtained in the probe brane
approximation for the fermionic part of the supergravity
ansatz. We will demand 
that the Killing spinors for the supergravity description lie in a
subbundle of the spin bundle spanned by the Killing spinors of the
probe brane description. We will further assume that the supergravity
Killing spinors are, up to multiplication by arbitrary functions, the
same as the probe brane Killing spinors given above. This second
assumption is the same
as saying that we assume the supergravity Killing spinors to be
simultaneous eigenspinors of the five projection operators of
\eqref{proj2} and \eqref{m2proj}\footnote{This second assumption may in fact
  be redundant. For all of the cases involving M5 branes studied in
  \cite{wrap} for which this was checked, orthogonality of the
  supergravity Killing spinors was 
  in fact implied by the Killing spinor equation. However, we have not
  checked this for the case in hand.}. 
For the bosonic part of the ansatz for the supergravity description,
we demand that the spacetime is a warped product of a timelike line
with a ten-manifold, so the metric is of the form of
\eqref{timelikeansatz}:
\bea
\dd s^{2} = -\Delta^{2} \dd t^{2} +\dd s^2(\mathcal{M}_{10}).
\eea
To complete our bosonic ansatz we demand that the magnetic flux
vanishes. This exhausts our assumptions in deriving the supergravity
description. 

Supersymmetry implies that the warp factor $\Delta$ and the frame on
$\mathcal{M}_{10}$ are 
independent of $t$, and furthermore that $\Delta$ is
related algebraically in the same way to the norms of the individual Killing
spinors, so the {\it a priori} arbitrary functions we allow for in the
Killing spinors must be the same. The remainder of the supersymmetry
conditions, which in the rest of this subsection we give for each
case in turn, will restrict the geometry of $\mathcal{M}_{10}$ and the
form of the electric flux. 

\subsubsection{Membranes wrapped in fourfolds: the supergravity description} 
In this case, the probe brane analysis revealed that the Killing
spinors preserved by the system define an $SU(5)$ structure, and are
algebraically identical to those for a membrane wrapped in a
five-fold. Given our assumptions for the supergravity description, we
have in fact already worked out the supersymmetry conditions for this
case; they are identical to those given for membranes wrapped in a fivefold.

In particular, any information regarding the original $SU(4)$
structure of the background is lost in the supergravity
description. As discussed in the introduction, we interpret this to
mean that the almost product structure of the spacetime is not
protected by supersymmetry in going to the supergravity regime.

\subsubsection{Membranes wrapped in threefolds: the supergravity description}
In this case, the probe brane Killing spinors define an $SU(3)\times
SU(2)$ structure. We take the supergravity Killing spinors to be
\bea\label{nmn}
\Delta^{1/2}\xi,\;\;\Delta^{1/2}\G^0\xi,\;\;\Delta^{1/2}\frac{1}{4}A^1_{ab}\G^{ab}\xi,\;\;\Delta^{1/2}\frac{1}{4}A^2_{ab}\G^{ab}\xi,
\eea
where now nothing is assumed about the frame $e^a$ on $\mathcal{M}_{10}$
beyond its $t$-independence. To derive the supersymmetry conditions
with these Killing spinors, we first impose that $\Delta^{1/2}\xi$ and
$\Delta^{1/2}\G^0\xi$ solve the Killing spinor equation with our
bosonic ansatz. Since together these spinors define an $SU(5)$
structure, they once again imply the supersymmetry conditions of
subsection \ref{kk}: $\mathcal{M}_{10}$ admits an $SU(5)$ structure,
and the torsion conditions and flux are given by
\bea 
\dd (e^0\w\mbox{Re}\Omega)&=&0,\nn
\dd\star J&=&0,\nn\label{mpak}
F&=&-\dd(e^0\w J).
\eea
Now we must impose the conditions implied by the existence of the
additional pair
$\eta_{(1)}=\Delta^{1/2}\frac{1}{4}A^1_{ab}\G^{ab}\xi,$
$\eta_{(2)}=\Delta^{1/2}\frac{1}{4}A^2_{ab}\G^{ab}\xi$. Observing that
$\G^0\eta_{(1)}=\eta_{(2)}$, we see that $\eta_{(1)}$ and $\eta_{(2)}$
collectively define a different $SU(5)$ structure. Therefore their
existence must imply the existence of a second solution of
\eqref{mpak}, but with different structure forms $(e^0)'$, $J'$, $\mbox{Re}\Omega'$. These
forms may be computed from the bilinears of $\eta_{(1)}$, and we find
\bea
(e^0)'&=&e^0,\nn
J'&=&J_{SU(3)}-J_{SU(2)},\nn
\mbox{Re}\Omega'&=&\mbox{Re}\Omega_{SU(3)}\w\mbox{Re}\Omega_{SU(2)}+\mbox{Im}\Omega_{SU(3)}\w\mbox{Im}\Omega_{SU(2)},
\eea
where we have defined
\bea
J_{SU(3)}&=&e^{12}+e^{34}+e^{56},\nn
\Omega_{SU(3)}&=&(e^1+ie^2)(e^3+ie^4)(e^5+ie^6),
\eea
and
\bea
J_{SU(2)}&=&e^{78}+e^{9\sharp},\nn
\Omega_{SU(2)}&=&(e^7+ie^8)(e^9+ie^{\sharp}).
\eea
By contrast, the $SU(5)$ forms defined by the Killing spinor
$\Delta^{1/2}\xi$ decompose under $SU(3)\times SU(2)$ according to
\bea
J&=&J_{SU(3)}+J_{SU(2)},\nn
\mbox{Re}\Omega&=&\mbox{Re}\Omega_{SU(3)}\w\mbox{Re}\Omega_{SU(2)}-\mbox{Im}\Omega_{SU(3)}\w\mbox{Im}\Omega_{SU(2)}. 
\eea 
Thus, on demanding that both the primed and the unprimed forms satisfy
\eqref{mpak}, we find the necessary and sufficient conditions for the
existence of the four Killing spinors \eqref{nmn} given our bosonic
ansatz. These conditions are
\bea
\dd
(\Delta\mbox{Re}\Omega_{SU(3)}\w\mbox{Re}\Omega_{SU(2)})&=&0,\nn
\dd(\D\mbox{Im}\Omega_{SU(3)}\w\mbox{Im}\Omega_{SU(2)})&=&0,\nn
\dd\star J_{SU(3)}=\dd \star J_{SU(2)}&=&0,\nn
\dd (\D J_{SU(2)})&=&0,\nn\label{opo}
F&=&-\dd (e^0\w J_{SU(3)}).
\eea 
In appendix \ref{mee}, we analyse these conditions in detail, and we
find that they may be considerably simplified. To state them, we first
define a quaternionic two-form $I_{SU(2)}$, in terms of
the unit quaternions $(i,j,k)$ and the invariant $SU(2)$ forms:
\bea
I_{SU(2)}=iJ_{SU(2)}+j\mbox{Re}\Omega_{SU(2)}+k\mbox{Im}\Omega_{SU(2)}.
\eea 
We find that \eqref{opo} implies that
\bea\label{emoo}
\dd (\D I_{SU(2)})=0.
\eea
Therefore the spacetime admits an almost product structure; if we
conformally rescale the metric along the $SU(2)$ directions according
to 
\bea
\dd s^2=-\D^2\dd t^2+\dd s^2(\mathcal{M}_{SU(3)})+\D^{-1}\dd
s^2(\mathcal{M}_{SU(2)}),
\eea
then \eqref{emoo} implies that $\mathcal{M}_{SU(2)}$ has $SU(2)$
holonomy, and the frame on it may be 
chosen to be independent of the coordinates, and the coordinate
differentials, of $\mathcal{M}_{SU(3)}$. The remaining conditions
constrain the $SU(3)$ structure on $\mathcal{M}_{SU(3)}$, together
with determining the flux. We find that they reduce to
\bea
\dd(e^0\w\Omega_{SU(3)}\w I_{SU(2)})&=&0,\nn
\dd\star J_{SU(3)}&=&0,\nn
F&=&-\dd(e^0\w J_{SU(3)}).
\eea
The formal similarity of these conditions with those for $SU(5)$ is
striking. From the first of these conditions we may read off all the
ways in which a probe M5 may be 
wrapped in the backreacted geometry, while preserving supersymmetry: on
the product of a SLAG three 
cycle in $\mathcal{M}_{SU(3)}$ (caibrated by either the real or
imaginary parts of $\Omega_{SU(3)}$) with a holomorphic curve (which could be
calibrated by $J_{SU(2)}$, $\mbox{Re}\Omega_{SU(2)}$ or
$\mbox{Im}\Omega_{SU(2)}$) in the $SU(2)$
manifold.

\subsubsection{Membranes wrapped in twofolds: the supergravity
  description} 
Now we turn to the supersymmetry conditions for membranes wrapped in
two-folds. The derivation proceeds in a very similar way to that for
threefolds. The most convenient way to obtain the conditions is to
observe that the $SU(2)$ structure defined by the Killing spinors in
this case is equivalent to a pair of $SU(3)\times SU(2)$ structures,
and to use the supersymmetry conditions for each. We will just state
the result. We find that the metric is given by
\bea
\dd s^2=-\D^2\dd t^2+\dd s^2(\mathcal{M}_{SU(2)})+\D^{-1}\dd
s^2(\mathbb{R}^6).
\eea
Thus in this case, the flatness of the overall transverse is protected
by supersymmetry to the same extent as it is for fivebranes. The
supersymmetry conditions fit the by-now familiar pattern; if $\D^{3/2}
\Lambda$ is an arbitrary closed three-form on the overall transverse
space, the supersymmetry conditions may be expressed as
\bea\label{331}
\dd (e^0\w\Omega_{SU(2)}\w\Lambda)&=&0,\\
\dd\star J_{SU(2)}&=&0,\\
F&=&-\dd(e^0\w J_{SU(2)}),
\eea
where here
\bea
J_{SU(2)} &=& e^{12} + e^{34}, \\
\Omega_{SU(2)} &=& (e^{1} + ie^{2} )(e^{3}+ie^{4}).
\eea  
Observe that \eqref{331} is equivalent to
$\dd(\D^{-1/2}\Omega_{SU(2)})=0$. Again, this is just a generalised calibration
condition for an M5 probe in the backreacted geometry. Observe however
that there is no condition of the form $\dd(e^0\w J_{SU(2)}\w
\Lambda)=0$. In an $SU(2)$ holonomy manifold, there is essentially no
distinction between holomorphic curves calibrated by $J$ or the real
or imaginary parts of $\Omega$. However because we have picked one of
the complex structures to calibrate the cycle wrapped by the
membranes, and then included backreaction, the symmetry of the complex
structures is broken, and so there is no $e^0\w J_{SU(2)}\w\Lambda$ generalised
calibration for probe fivebranes.

\section{The $AdS$ limits of wrapped membranes}
In this section, we study the $AdS$ limits, and associated
supersymmetry conditions, of the supergravity description of
membranes wrapped in threefolds and twofolds. As in section \ref{rrr}, this involves making a suitable
ansatz for the warp factor and frame, picking the $AdS$ radial
direction out of the ten Riemannian dimensions, and imposing vanishing
of the flux components not containing the $AdS$ volume form as a
factor. Picking out the $AdS$ radial direction is now somewhat less
trivial, because the wrapped brane metrics in these cases are
\bea
\dd s^2=-\D^2\dd t^2+\dd s^2(\mathcal{M}_{10-p})+\D^{-1}\dd
s^2(\mathcal{N}_p),
\eea
where $\mathcal{N}_p$ is an $SU(2)$ holonomy manifold for membranes
wrapped in a threefold and $\mathcal{N}_p=\mathbb{R}^6$ for a
twofold. For the case of a threefold, we will restrict attention to
the special case $\mathcal{N}_p=\mathbb{R}^4$ henceforth. It would be
interesting to know if there are other choices for the $SU(2)$
manifold that admit an $AdS$ limit, but we think this is unlikely and
we will not pursue this question here. Thus the
wrapped brane metrics we study are
\bea
\dd s^2=-\D^2\dd t^2+\dd s^2(\mathcal{M}_{9-q})+\D^{-1}(\dd R^2+R^2\dd s^2(S^q)),
\eea 
where $q=3$ for threefolds and $q=5$ for twofolds. Generically, the
$AdS$ radial direction will lie partly in $\mathcal{M}_{9-q}$ and
partly in the overall transverse space. It cannot lie entirely in
$\mathcal{M}_{9-q}$, because then explicit dependence on the $AdS$
radial coordinate will enter through the $\D^{-1}$ warp factor of the
overall transverse space, in contradiction of our assumption of a
warped $AdS$ product. However we will see that for threefolds (but not
for twofolds) the $AdS$ radial direction can lie entirely in the
overall transverse space. This non-generic case will be discussed
seperately below, but here we will focus on describing our limiting procedure in
the generic case, where the $AdS$ radial direction lies partly in
$\mathcal{M}_{9-q}$ and partly in the overall transverse space. Our
treatment very closely follows that of \cite{wrap}. We
will assume that the part of the $AdS$ radial direction lying in the overall
transverse lies entirely along the radial direction of the overall
transverse space. We emphasise that this is indeed an assumption, and
in contrast to the $SU(5)$ case, we are no longer gauranteed that we
are taking the most general $AdS$ limit of the wrapped brane
configurations (though we believe that in fact we are). 

We may extract the $AdS$ radial direction by performing a frame
rotation. Defining, as in section \ref{rrr}, 
\bea
\D=e^{-mr}\l^{-1/2},
\eea
where $r$ is the $AdS$ radial coordinate, we write, for some one-form
$\hat{u}$ lying entirely in $\mathcal{M}_{9-q}$ and for
$\hat{v}=\Delta^{-1/2}\dd R$,
\bea
\l^{-1/2}\dd r=\sin\theta\hat{u}+\cos\theta\hat{v},
\eea
where we assume that the rotation angle $\theta$ is independent of
$r$. We also define the orthogonal combination
\bea
\hat{\r}=\cos\theta\hat{u}-\sin\theta\hat{v}.
\eea
Inverting these expressions to give $\hat{v}$ in terms of the new
frame, then imposing closure of $\dd R$, we find that
\bea
\hat{\r}=\frac{2\l^{1/4}}{m\sin\theta}\dd(\l^{-3/4}\cos\theta).
\eea 
Defining a new coordinate $\rho=\l^{-3/4}\cos\theta$, we get
\bea
\hat{\r}&=&\frac{2\l^{1/4}}{m\sss}\dd\r,\nn
\hat{u}&=&\l^{-1/2}\sss\dd r+\frac{2\l\r\dd \r}{m\sss},\nn
R&=&-\frac{2}{m}\r e^{-mr/2}.
\eea
Therefore, the spacetime metric in this $AdS$ limit becomes
\bea
\dd s^2=\frac{1}{\l m^2}\Big(\dd s^2(AdS_2)+4\l^{3/2}\Big[\frac{\dd
  \r^2}{1-\l^{3/2}\r^2}+\r^2\dd s^2(S^q)\Big]\Big)+\dd
s^2(\mathcal{M}_{8-q}),
\eea
where $\dd s^2(\mathcal{M}_{8-q})$ is defined by
\bea
\dd s^2(\mathcal{M}_{9-q})=\dd
s^2(\mathcal{M}_{8-q})+\hat{u}\otimes\hat{u}.
\eea
In taking this $AdS$ limit, we are picking out preferred vectors
(associated to the additional $AdS$ Killing spinors) in both
$\mathcal{M}_{9-q}$ and the overall transverse space. This will reduce
the structure group of the wrapped brane spacetime: from $SU(3)\times
SU(2)$ to an $SU(2)$ subgroup of the $SU(3)$ factor for threefolds,
and from $SU(2)$ to the identity for twofolds. To complete the $AdS$
limit, we impose vanishing of the flux components not containing a
factor of the $AdS$ volume form.

\subsection{The $AdS$ limit of membranes on threefolds}
As we have mentioned, in addition to the more generic limiting procedure
discussed above, the supergravity description for membranes wrapped in
a threefold admits a special $AdS$ 
limit, where the $AdS$ radial direction is assumed to lie entirely in
the overall transverse space. We will first discuss this special case,
before moving on to the more generic one.

\subsubsection{$AdS$ radial direction from the overall transverse
  space}
Demanding that the $AdS$ radial coordinate lies entirely in the
overall transverse space implies that $\l$ is constant (and may be set
to unity by rescaling) and that $\partial/\partial r$ lies along the
radial direction of the overall transverse space. Since in this case no
preferred vector associated to a Killing spinor is picked out on the
$SU(3)$ manifold, the structure group of the $AdS$ limit is reduced to
$SU(3)$ here. The metric is given by
\bea
\dd s^2=\frac{1}{m^2}[\dd s^2(AdS_2)+4\dd s^2(S^3)]+\dd
s^2(\mathcal{M}_{SU(3)}).
\eea
Then the supersymmetry conditions imply that
\bea
\dd J_{SU(3)}=\dd \Omega_{SU(3)}=0,
\eea
and hence the spacetime is the direct product $AdS_2\times S^3\times
CY_3$. The flux is given by $F=-\mbox{Vol}_{AdS_2}\w
J_{SU(3)}/m$. This solution is of course well known. For fivebranes
wrapped on K\"{a}hler four-cycles, it is also possible to take a
non-generic $AdS$ limit of the wrapped brane supersymmetry conditions,
in exactly the same way as is done here \cite{wrap}. The non-generic fivebrane
$AdS_3$ limit turns out to be the analytic continuation of the
non-generic membrane $AdS_2$ limit. 

\subsection{Generic case}
The $AdS$ supersymmetry conditions in the more generic case, where the
$AdS$ radial direction is assumed to lie partly in the $SU(3)$
structure manifold and partly along the radial direction of the
overall transverse space, are worked out in detail in appendix
\ref{FF}. The metric is given by
\bea
\dd s^2&=&\frac{1}{\l m^2}\Big(\dd s^2(AdS_2)+4\l^{3/2}\Big[\frac{\dd
  \r^2}{1-\l^{3/2}\r^2}+\r^2\dd s^2(S^3)\Big]\Big)\nn&&+\dd
s^2(\mathcal{M}_{SU(2)})+w\otimes w,
\eea
where $\mathcal{M}_{SU(2)}$ admits an $SU(2)$ structure. If we define
the $SU(2)$ structure forms
\bea
J^1&=&e^{12}+e^{34},\nn
J^2&=&e^{14}+e^{23},\nn
J^3&=&e^{13}-e^{24},
\eea
the conditions on the intrinsic torsion are
\bea\label{414}
\dd \left(\l^{-1/2}\sss J^2\right)&=&0,\\\label{415}
\dd \left(\l^{-1/2}\sss J^3\right)&=&0,\\\label{416}
\dd (\l^{-1/2}J^1+\l^{1/4}\r\hat{w}\w\hat{\r})&=&0,\\\label{417}
J^3\w
\dd\left(\frac{\l^{1/2}}{\sss}\hat{w}\right)&=&J^2\w\dd\left(\frac{1}{\l^{1/4}\r\sss}\hat{\r}\right),\\\label{419}
J^2\w
\dd\left(\frac{\l^{1/2}}{\sss}\hat{w}\right)&=&-J^3\w\dd\left(\frac{1}{\l^{1/4}\r\sss}\hat{\r}\right).
\eea
The flux is given by
\bea\label{420}
F=\frac{1}{m^2}\mbox{Vol}_{AdS_2}\w\left[\dd\left(\l^{-1}\sss\hat{w}\right)-m(\l^{-1/2}J^1+\l^{1/4}\r\hat{w}\w\hat{\r})\right].
\eea
As discussed in the introduction, in the generic
case also, the supersymmetry conditions we obtain are the analytic
continuation of the conditions of \cite{wrap} for the generic $AdS_3$
limit of M5s
wrapped on a K\"{a}hler four-cycle in a threefold. For the M5s, the
isometry of the generic $AdS_3$ limit is $SO(3)\times U(1)$; for the
membranes, the equations given above imply that the isometry of the
generic $AdS_2$ limit is $SO(4)\times U(1)$, with the $SO(4)$ coming
from the sphere and the $U(1)$ generated by $\hat{w}$. We may also analytically continue so that the $U(1)$ isometry
becomes the timelike direction, and so obtain supersymmetric M-theory
spacetimes with $SO(4)\times SO(3)$ isometry. We identify the
spacetimes satisfying these supersymmetry conditions as 1/4 BPS
bubbling solutions in M-theory.

\subsection{The $AdS$ limit of membranes on twofolds}
Now we will give the conditions we derive for the $AdS$ limit of
membranes on a twofold. Observe in that in this case, $\dd(\D^{-1/2}
\Omega_{SU(2)})=0$, it is not possible to take a non-generic $AdS$
limit, with the $AdS$ radial direction lying entirely in the overall
transverse space. The derivation of the supersymmetry conditions is
very similar to that for threefolds in the generic case, and technical
details are omitted. The metric is given by
\bea
\dd s^2&=&\frac{1}{\l m^2}\Big(\dd s^2(AdS_2)+4\l^{3/2}\Big[\frac{\dd
  \r^2}{1-\l^{3/2}\r^2}+\r^2\dd s^2(S^5)\Big]\Big)\nn&&+e^1\otimes
e^1+e^2\otimes e^2+e^3\otimes e^3,
\eea
with $\hat{w}=e^3$. The torsion conditions we derive are
\bea
\dd\left(\l^{-1/4}\sss
  e^1\right)&=&-\frac{m\l^{1/4}}{2}\left(\l^{3/4}\r e^1\w
  \hat{\r}+e^{23}\right),\\
\dd\left(\l^{-1/4}\sss
  e^2\right)&=&-\frac{m\l^{1/4}}{2}\left(\l^{3/4}\r e^2\w
  \hat{\r}-e^{13}\right),\\
\dd\left(\frac{\l^{1/2}}{\sss
  }e^3\right)&=&-\frac{2m\l}{1-\l^{3/2}\r^2}e^{12}-\frac{3\l^{1/4}\r}{2(1-\l^{3/2}\r^2)^{3/2}}(\partial_{\hat{\r}}\l
e^{12}\nn
&&-\partial_2\l e^1\w\hat{\r}+\partial_1\l e^2\w\hat{\r}).
\eea
The flux is then given by
\bea
F&=&\frac{1}{m^2}\mbox{Vol}_{AdS_2}\w\left[\dd\left(\l^{-1}\sss
    e^3\right)-m\l^{-1/2}\left(e^{12}+\l^{3/4}\r
    e^3\w\hat{\r}\right)\right].
\eea
Upon analytic continuation, these give precisely the conditions of LLM
\cite{41}, expressed in a form similar to that of \cite{wrap}.

\section{Conclusions}
In this work, we have performed a systematic study of the geometry of
wrapped brane configurations admitting $AdS_2$ limits in eleven
dimensional supergravity. We have found that the backreaction of
wrapped membranes (in threefolds or fourfolds) on their overall
transverse space is less restricted by the kinematics of the Killing
spinor equation than that hitherto observed for fivebranes on any
cycle. The reason for this can ultimately be traced back to the
Killing spinors. For the cycles wrapped by fivebranes studied in
\cite{wrap}, it was found that the Killing spinors preserved in the
presence of the brane are isomorphic to vectors of $Spin(7)$ (recall
that the isotropy group of a null spinor in eleven dimensions is
$(Spin(7)\ltimes\mbb^8)\times\mbb$; for the wrapped fivebrane
configurations of \cite{wrap}, the Killing spinors are null). These
vectors essentially lock the frame on the overall transverse space in
passing from the probe brane to the supergravity description, and it
is their presence which ultimately restricts the gravitational effect
of the fivebrane on the overall transverse space. By contrast, the
Killing spinors preserved by probe membranes are isomorphic to zero or
two-forms of $SU(5)$; the vectors are projected out by the
kappa-symmetry projections. Thus (for fourfolds and threefolds) one
cannot define preferred vectors on the overall transverse space
associated to the supersymmetries, and this is ultimately the
kinematical origin of the effect we have observed. 

We have seen how the supersymmetry conditions for wrapped
membranes in Calabi-Yau $n$-folds may be expressed in a simple
universal form. We have also made manifest the link, via analytic
continuation, between the $AdS$ limits of the supergravity description
of membranes and fivebranes wrapped in threefolds or twofolds, and 1/4
and 1/2 BPS bubbling solutions in M-theory.

It will be interesting to examine these 1/4-BPS bubbling geometries in
more detail, and in particular, to try to find some explicit solutions
of the supersymmetry conditions (assuming, of course, that
some exist) together with their field theory duals. In the 1/2-BPS case, different boundary conditions must
be imposed in the different ($AdS_2$, $AdS_5$ or bubbling) branches in
order to get regular solutions; it will be interesting to see if
something similar applies here. 

It will also be interesting to apply the techniques of \cite{wrap} to
the study of fivebranes wrapping four-cycles in eight-manifolds. This
case is particularly rich, and there are many possibilities to
consider. One can include membranes, intersecting the fivebranes
in a string, and extended in the directions transverse to the
eight-manifold. The new explicit $AdS_3$ solutions of \cite{46},
\cite{47} come from this sector of M-theory. A general analysis of the
different possible cases is under investigation \cite{50}.

\section{Acknowledgements}
OC is supported by EPSRC. We are grateful to Pau Figueras for initial
collaboration on this project, and to Jerome Gauntlett for useful
discussions and comments.

\appendix

\section{Branes wrapping cycles in five-folds: technical details}
In this appendix, we will give the technical details of the derivation
of the supersymmetry conditions for the near-horizon limits of branes
wrapping cycles in $SU(5)$ manifolds.

\subsection{The $AdS$ limit of fivebranes on SLAG five-cycles}
To derive the $AdS$ supersymmetry conditions, we first decompose the
$SU(5)$ modules of the wrapped brane structure group into modules of
the $SU(4)$ structure group of the near-horizon limit, under which the
metric decomposes according to
\bea
ds^2(\mathcal{M}_{10})=ds^2(\mathcal{M}_8)+e^9\otimes e^9+\l^{-1}\dd
r^2,
\eea
with $\mathcal{M}_8$ admitting an $SU(4)$ structure. We start with the
flux term in the $\mathbf{75}$ of $SU(5)$. Under $SU(4)$, this
decomposes as 
\bea
\mathbf{75}\rightarrow\mathbf{20}+\mathbf{15}+(\mathbf{20}+\bar{\mathbf{20}}). 
\eea
The flux term in the $\mathbf{75}$ of $SU(5)$ thus decomposes
according to
\bea
F^{\mathbf{75}} &=& F_1^{\mathbf{20}}+F_2^{\mathbf{15}}\w
(J_{SU(4)}-2e^{9\sharp})+F_3^{(\mathbf{20}+\bar{\mathbf{20}})}\w
e^9 \nn &-& J_{SU(4)}\cdot
F_3^{(\mathbf{20}+\bar{\mathbf{20}})}\w e^{\sharp},
\eea
where for an $n$-from $\Lambda$, we have defined
$J\cdot\Lambda_{i_1...i_n}=nJ_{[i_1}^{\;\;j}\Lambda_{|j|i_2...i_n]}$,
and we have used $J_{SU(5)} \lrcorner F^{\mathbf{75}} = 0$.

Next, we decompose the $SU(5)$ torsion modules into modules of
$SU(4)$. Since
$(\mathbf{10}+\bar{\mathbf{10}})\rightarrow(\mathbf{6}+\bar{\mathbf{6}})+(\mathbf{4}+\bar{\mathbf{4}})$,
$W_1$ decomposes according to
\bea
W_1^{(\mathbf{10}+\bar{\mathbf{10}})} 
= A_1^{(\mathbf{6}+\bar{\mathbf{6}})}+A_2^{(\mathbf{4}+\bar{\mathbf{4}})}\w e^9  + J_{SU(4)}\cdot A_2^{(\mathbf{4}+\bar{\mathbf{4}})}\w
e^{\sharp}.
\eea
For $W_2$, since
$(\mathbf{40}+\bar{\mathbf{40}})\rightarrow(\mathbf{20}+\bar{\mathbf{20}})+(\mathbf{10}+\bar{\mathbf{10}})+(\mathbf{6}+\bar{\mathbf{6}})+(\mathbf{4}+\bar{\mathbf{4}})$,
we find that
\bea
W_2^{(\mathbf{40}+\bar{\mathbf{40}})}&=&B_1^{(\mathbf{10}+\bar{\mathbf{10}})}+B_2^{(\mathbf{6}+\bar{\mathbf{6}})}\w(J_{SU(4)}-2e^{9\sharp})+(B_3^{(\mathbf{4}+\bar{\mathbf{4}})}\lrcorner
\mbox{Im}\Omega_{SU(4)} +B_4^{(\mathbf{20}+\bar{\mathbf{20}})})\w e^9\nn&+& (-B_3^{(\mathbf{4}+\bar{\mathbf{4}})}\lrcorner
\mbox{Re}\Omega_{SU(4)}+J_{SU(4)}\cdot B_4^{(\mathbf{20}+\bar{\mathbf{20}})})\w
e^{\sharp}.
\eea
For $W_3$, since
$(\mathbf{45}+\bar{\mathbf{45}})\rightarrow(\mathbf{20}+\bar{\mathbf{20}})+(\mathbf{6}+\bar{\mathbf{6}})+(\mathbf{4}+\bar{\mathbf{4}})+\mathbf{15}+\mathbf{15}'$,  
we have the decomposition
\bea
W_3^{(\mathbf{45}+\bar{\mathbf{45}})}&=&C_1^{(\mathbf{20}+\bar{\mathbf{20}})}+C_2^{(\mathbf{4}+\bar{\mathbf{4}})}\w(J_{SU(4)}-3e^{9\sharp})+(C_3^{(\mathbf{6}+\bar{\mathbf{6}})}+C_4^{\mathbf{15}})\w
  e^9\nn&+&\Big(-\frac{1}{2}J_{SU(4)}\cdot C_3^{(\mathbf{6}+\bar{\mathbf{6}})}+C_5^{\mathbf{15}'}\Big)\w
  e^{\sharp}.
\eea
The modules $W_4$ and $W_5$ decompose as vectors,
$(\mathbf{5}+\bar{\mathbf{5}})\rightarrow(\mathbf{4}+\bar{\mathbf{4}})+(\mathbf{1}+\bar{\mathbf{1}})$:
\bea
\mathcal{W}_4&=&D_1^{(\mathbf{4}+\bar{\mathbf{4}})}+D_2e^9+D_3e^{\sharp},\nn
\mathcal{W}_5&=&E_1^{(\mathbf{4}+\bar{\mathbf{4}})}+E_2e^9+E_3e^{\sharp}.
\eea
Now to obtain the flux and torsion conditions in the $AdS$ limit, we
simply impose vanishing of the magnetic flux components along
$e^{\sharp}$, the vanishing of electric flux components not containing a
factor proportional to the $AdS$ volume form, and also decompose both
sides of equation \eqref{213} under $SU(4)$. Defining $Z=\dd
\log\Delta+\mathcal{W}_4$, imposing vanishing of the magnetic
flux along the $AdS$ radial direction, we find that
\bea
B_2^{(\mathbf{6}+\bar{\mathbf{6}})}&=&\frac{1}{3}A_1^{(\mathbf{6}+\bar{\mathbf{6}})},\nn
A_2^{(\mathbf{4}+\bar{\mathbf{4}})}&=&0,\nn
B_3^{(\mathbf{4}+\bar{\mathbf{4}})}&=&-\frac{1}{2}Z^{(\mathbf{4}+\bar{\mathbf{4}})}.
\eea
The surviving magnetic flux is then given by
\bea
F_{mag}&=&-\frac{1}{2}(B_1^{(\mathbf{10}+\bar{\mathbf{10}})}+3B_2^{(\mathbf{6}+\bar{\mathbf{6}})}\w
J_{SU(4)})+\frac{1}{4}(Z_9\mbox{Im}\Omega_{SU(4)}+Z_{\sharp}\mbox{Re}\Omega_{SU(4)})\nn&-&(B_4^{(\mathbf{20}+\bar{\mathbf{20}})}+B_3^{(\mathbf{4}+\bar{\mathbf{4}})}\lrcorner\mbox{Im}\Omega_{SU(4)})\w e^9.
\eea
Next we decompose equations \eqref{213} under $SU(4)$. The equation
for $\dd J$ implies that
\bea\label{A9}
\dd J_{SU(4)}&=&\Big(C_2^{(\mathbf{4}+\bar{\mathbf{4}})}+\frac{1}{4}D_1^{(\mathbf{4}+\bar{\mathbf{4}})}\Big)\w
J_{SU(4)}+C_1^{(\mathbf{20}+\bar{\mathbf{20}})}\nn
&+&\Big(\frac{3}{8}B_2^{(\mathbf{6}+\bar{\mathbf{6}})}\lrcorner\mbox{Im}\Omega_{SU(4)}+C_3^{(\mathbf{6}+\bar{\mathbf{6}})}+C_4^{\mathbf{15}}+\frac{1}{4}D_2J_{SU(4)}\Big)\w
e^9,\\\label{A10}
\l^{1/2}\dd
(\l^{-1/2}e^9)&=&\frac{3}{8}B_2^{(\mathbf{6}+\bar{\mathbf{6}})}\lrcorner\mbox{Re}\Omega_{SU(4)}
-\frac{1}{2}J_{SU(4)}\cdot
C_3^{(\mathbf{6}+\bar{\mathbf{6}})}+C_5^{\mathbf{15}'}\nn&+&\frac{1}{4}D_3J_{SU(4)}+\Big(\frac{1}{4}D_1^{(\mathbf{4}+\bar{\mathbf{4}})}-3C_2^{(\mathbf{4}+\bar{\mathbf{4}})}\Big)\w
e^9.
\eea
From the equation for $\dd \mbox{Re}\Omega$, we get
\bea
&&\l^{1/2}\dd(\l^{-1/2}\mbox{Im}\Omega_{SU(4)})=\Big(B_3^{(\mathbf{4}+\bar{\mathbf{4}})}\lrcorner\mbox{Re}\Omega_{SU(4)}-J_{SU(4)}\cdot
B_4^{(\mathbf{20}+\bar{\mathbf{20}})}\Big)\w J_{SU(4)}\nn&&-\frac{1}{8}\mbox{Im}\Omega_{SU(4)}\w
E_1^{(\mathbf{4}+\bar{\mathbf{4}})}-\Big[B_1^{(\mathbf{10}+\bar{\mathbf{10}})}+\frac{1}{8}\Big(E_3\mbox{Re}\Omega_{SU(4)}+E_2\mbox{Im}\Omega_{SU(4)}\Big)\Big]\w
e^9,\nn\label{A11}
\eea
\bea
\dd(\mbox{Re}\Omega_{SU(4)}\w
e^9)&=&\frac{3}{2}B_2^{(\mathbf{6}+\bar{\mathbf{6}})}\w
J^2_{SU(4)}+\Big[\Big(B_3^{(\mathbf{4}+\bar{\mathbf{4}})}\lrcorner\mbox{Im}\Omega_{SU(4)}+B_4^{(\mathbf{20}+\bar{\mathbf{20}})}\Big)\w
J_{SU(4)}\nn\label{A13}&&
-\frac{1}{8}\mbox{Re}\Omega_{SU(4)}\w
E_1^{(\mathbf{4}+\bar{\mathbf{4}})}\Big]\w e^9.
\eea
Next, imposing that the electric flux contains a factor proportional
to the $AdS$ volume form, we get
\bea\label{A12}
\dd (\l^{-1/2}J_{SU(4)})&=&0,\\
F_{elec}&=&\frac{1}{m^2}\mbox{Vol}_{AdS_2}\w[\dd
(\l^{-1}e^9)-m\l^{-1/2}J_{SU(4)}].
\eea
Finally we must impose the $SU(5)$ torsion condition,
$\mathcal{W}_5=8\dd\log\Delta$. 

To make further progress, we first compare equation \eqref{A9} with
\eqref{A12}. This implies the following torsion conditions
\bea
C_1^{(\mathbf{20}+\bar{\mathbf{20}})}=C_4^{\mathbf{15}}&=&0,\nn
B_2^{(\mathbf{6}+\bar{\mathbf{6}})}\lrcorner\mbox{Im}\Omega_{SU(4)}&=&-\frac{8}{3}C_3^{(\mathbf{6}+\bar{\mathbf{6}})},\nn
C_2^{(\mathbf{4}+\bar{\mathbf{4}})}+\frac{1}{4}D_1^{(\mathbf{4}+\bar{\mathbf{4}})}&=&\frac{1}{2}\tilde{\dd}\log\l,\nn
D_2&=&2\partial_9\log\l,
\eea
where here $\tilde{\dd}$ denotes the exterior derivative restricted to
$\mathcal{M}_8$. At this point, using the algebraic restrictions we
have derived on the torsion, we may eliminate all vector and singlet
modules in favour of $D_1^{(\mathbf{4}+\bar{\mathbf{4}})}$, $D_3$, and
derivatives of $\l$. Upon doing this, we find that \eqref{A13} is in
fact implied by equations \eqref{A10} and \eqref{A11}. We may thus
summarise the supersymmetry conditions as
\bea
\dd(\l^{-1/2}J)&=&0,\nn
\dd
e^9&=&\frac{3}{4}B_2^{(\mathbf{6}+\bar{\mathbf{6}})}\lrcorner\mbox{Re}\Omega_{SU(4)}+C_5^{\mathbf{15}'}+\frac{1}{4}D_3J+(D_1^{(\mathbf{4}+\bar{\mathbf{4}})}-\tilde{\dd}\log\l)\w e^9,\nn
\dd \mbox{Im}\Omega_{SU(4)}&=&-J_{SU(4)}\cdot
B_4^{(\mathbf{20}+\bar{\mathbf{20}})}\w
J_{SU(4)}+\mbox{Im}\Omega_{SU(4)}\w\Big(\frac{5}{4}\tilde{\dd}\log\l-\frac{1}{2}D_1^{(\mathbf{4}+\bar{\mathbf{4}})}\Big)\nn&+&\Big[\l^{1/2}m\mbox{Re}\Omega_{SU(4)}+\partial_9\log\l\mbox{Im}\Omega_{SU(4)}-B_1^{(\mathbf{10}+\bar{\mathbf{10}})}\Big]\w
e^9,\nn
F&=&\frac{1}{m^2}\mbox{Vol}_{AdS_2}\w[\dd
(\l^{-1}e^9)-m\l^{-1/2}J_{SU(4)}]+F_1^{\mathbf{20}}-\frac{1}{2}B_1^{(\mathbf{10}+\bar{\mathbf{10}})}\nn&-&\frac{3}{2}B_2^{(\mathbf{6}+\bar{\mathbf{6}})}\w
J_{SU(4)}+\frac{3}{8}\partial_9\log\l\mbox{Im}\Omega_{SU(4)}+\frac{1}{4}(D_3-m\l^{1/2})\mbox{Re}\Omega_{SU(4)}\nn&-&\Big[B_4^{(\mathbf{20}+\bar{\mathbf{20}})}+\frac{1}{4}(\tilde{\dd}\log\l-2D_1^{(\mathbf{4}+\bar{\mathbf{4}})})\lrcorner\mbox{Im}\Omega_{SU(4)}\Big]\w
e^9.
\eea
Relabelling the torsion modules we obtain the expressions quoted in
the main text.

\subsection{The $AdS_2$ limit of membranes on holomorphic curves}
We will now give further technical details of the derivation of the
$AdS$ limit of the supersymmetry conditions for membranes on
holomorphic curves in five-folds, \eqref{electricsu5}. As
discussed in the main text, there are two equivalent ways in which
these conditions may be arrived at; though as it is quicker to
obtain the result by setting the magnetic fluxes to zero in
\eqref{phtoo}, this is what will be presented here. The metric in the
$AdS$ limit is
\bea
\dd s^2=\frac{1}{\l m^2}\dd s^2(AdS_2)+\dd
s^2(\mathcal{M}_8)+e^9\otimes e^9,
\eea
with $\mathcal{M}_8$ admitting an $SU(4)$ structure. Throughout this
subsection, all forms and modules are those of $SU(4)$.

Thus, setting the magnetic fluxes to zero in \eqref{phtoo}, the
conditions \eqref{phto} reduce to
\bea\label{45}
\partial_9\l&=&0,\\\label{46}
\dd(\l^{1/2}e^9)&=&\frac{1}{4}m\l J+\mathcal{W}^{\mathbf{15}},\\\label{42}
\dd (\l^{-1/2}J)&=&0,\\\label{44}
\dd (\l^{-1}\mbox{Im}\Omega)&=&m\l^{-1/2}e^9\w\mbox{Re}\Omega,
\eea
where $\mathcal{W}^{\mathbf{15}}$ is a two-form in the adjoint of
$SU(4)$ which is unfixed by the supersymmetry conditions.
  
To make progress in solving these conditions, let us introduce
coordinates $z,\;x$ such that 
\bea
e^9=\frac{A}{m}(\dd z+\s),
\eea
where $\s$ is a one-form defined on the base $SU(4)$ manifold with
coordinates $x$, and {\it a priori} $A$ and $\s$ depend on
$z,\;x$. Now, we observe that (\ref{45}) together with the 9 component
of (\ref{42}) implies that
\bea
\partial_z\l=\partial_z J=0.
\eea
Next, the 9 component of (\ref{46}) gives
\bea
\partial_z(\l^{1/2}A\s)=\tilde{\dd}(\l^{1/2}A),
\eea
or
\bea
\s=\frac{1}{\l^{1/2}A}\Big(\tilde{\dd}\int^z\l^{1/2}A\dd z'+\s_0(x)\Big).
\eea
By choosing a new coordinate, we may always take
$A=\l^{-1/2}$, $\s=\s(x)$; explicitly, we choose
\bea
z'=\int^z\l^{1/2}A\dd z''.
\eea
Then $e^9=m^{-1}\l^{-1/2}(\dd z'+\s_0(x))=m^{-1}e^{9'}$; dropping the primes and subscripts,
this is the gauge in which we will work henceforth.

At this point, it is convenient to conformally rescale the base space,
$g_8=m^{-2}\l^{1/2}\hat{g}_8,\;J=m^{-2}\l^{1/2}\hat{J},\;
\Omega=m^{-4}\l\hat{\Omega}$. Henceforth we will work only
  with the rescaled base metric, and drop the hats. Then
  (\ref{44}) becomes 
\bea\label{47}
\dd\mbox{Im}\Omega=\l^{1/2}\mbox{Re}\Omega\w e^9.
\eea
If we define
\bea
\Lambda_{ij}=(\partial_ze^i)_j,
\eea
then the $z$-independence of $J$ together with the 9 component
of (\ref{47}) are equivalent to
\bea
\Lambda_{ij}=-\frac{1}{4}J_{ij}+\Lambda^{\mathbf{15}}_{ij}.
\eea
We may always eliminate the $\Lambda^{\mathbf{15}}$ term by performing a
$z$-dependent $SU(4)$ rotation of the frame on the base. Thus we may
solve for the $z$-dependence of the frame for the base space, according to
\bea
e^1&=&-\sin\frac{z}{4}\tilde{e}^2(x)+\cos\frac{z}{4}\tilde{e}^1(x),\nn 
e^2&=&\cos\frac{z}{4}\tilde{e}^2(x)+\sin\frac{z}{4}\tilde{e}^1(x),
\eea
and similarly for the other pairs of basis one-forms. This rotation to
the tilded frame leaves the metric and $J$ invariant, but
shifts $\Omega$ by a phase:
\bea
\Omega(z,x)=e^{iz}\tilde{\Omega}(x).
\eea
Then the remaining content of (\ref{47}) may be expressed as
\bea
\dd\tilde{\Omega}=i\s\w\tilde{\Omega}.
\eea
This, together with $dJ=0$, implies that the conformally
rescaled base admits a K\"{a}hler metric, with Ricci form 
\bea
\mathcal{R}=d\s.
\eea
Observe that this is consistent with the absence of $(2,0)+(0,2)$
forms in \eqref{46}. The final remaining condition comes from the
singlet of (\ref{46}). This is equivalent to the condition
\bea
R=2\l^{3/2},
\eea
where $R$ is the scalar curvature of the base. These conditions,
summarised in the main text, are precisely those of \cite{nakwoo}.

\section{Membranes on $SU(n)$ manifolds: technical details}\label{mee}
In this appendix, we will give more technical details of the
supersymmetry conditions for membranes wrapped in threefolds. The
derivation for twofolds is similar and is omitted. 

In the main text, a set of supersymmetry conditions for the
supergravity description of membranes wrapped in threefolds was
derived. Manipulating these conditions into the final form subsequently
quoted is a useful exercise in $SU(3)\times SU(2)$ structures. We
start by conformally rescaling the $SU(2)$ directions according to 
\bea
\dd s^2=-\D^2\dd t^2+\dd s^2(\mathcal{M}_{SU(3)})+\D^{-1}\dd
s^2(\mathcal{M}_{SU(2)}).
\eea
Then in terms of the rescaled forms (which we use throughout this
section) the equations of the main text become
\bea\label{b2}
\dd
(\mbox{Re}\Omega_{SU(3)}\w\mbox{Re}\Omega_{SU(2)})&=&0,\\\label{b3}
\dd(\mbox{Im}\Omega_{SU(3)}\w\mbox{Im}\Omega_{SU(2)})&=&0,\\\label{b4}
\dd J_{SU(2)}&=&0,\\\label{b5}
J_{SU(2)}\w\dd\mbox{Vol}_{SU(3)}&=&0,\\\label{b6}
\mbox{Vol}_{SU(2)}\w\dd(\D^{-1}J_{SU(3)}^2)&=&0.
\eea 
To analyse these equations, we introduce the following notation. Let
upper-case letters $A,B,..=1,...,6$ denote the $SU(3)$ directions, and
let lower-case letters $a,b=7,...,\sharp$ denote the $SU(2)$
directions. Let $\tilde{\dd}$ denote the exterior derivative
restricted to the 1,...,6 directions, and let $\hat{\dd}$ denote the
exterior derivative restricted to the $7,...,\sharp$
directions. We say a form is a $(p;q)$ form if it has $p$ indices
along the $SU(3)$ directions and $q$ indices along the $SU(2)$
directions. Furthermore, let us define
\bea
\dd e^A&=&\frac{1}{2}U^A_{ab}e^a\w e^b+V^{AB}_ae^a\w
e^B+\tilde{\dd}e^A,\\
\dd e^a&=&\frac{1}{2}X^a_{AB}e^A\w e^B+Y^{ab}_Ae^A\w
e^b+\hat{\dd}e^a.
\eea
Now we begin the analysis. First, observe that the only torsion module
contained in \eqref{b6} is (in standard notation) the $\mathcal{W}_4$
module of the $SU(3)$ structure on $\mathcal{M}_6$. Furthermore, this
module appears nowhere else, so \eqref{b6} is completely independent
of the other equations. Next we look at \eqref{b4}. We may re-express
this as
\bea
J_{SU(2)ab}X^a\w e^b+J_{SU(2)ab}Y^{ac}\w e^c\w
e^b+\hat{\dd}J_{SU(2)}=0.
\eea
Since the first of the terms is a (2;1) form, the second a (1;2) form
and the third a (0;3) form, they must all vanish seperately. Thus
\bea
\hat{\dd}J_{SU(2)}&=&0,\\
X^a&=&0,\\\label{b7}
J_{SU(2)c[a}Y_{b]}^{\;c}&=&0.
\eea
Observe that on contracting the third of these equations with
$J_{SU(2)ab}$ we find that $Y^a_{\;\;a}=0$. Next look at \eqref{b5}. This is equivalent to 
\bea
J_{SU(2)}\w U^A&=&0,\\\label{b9}
V^A_{\;\;\;A}&=&0.
\eea
Equations \eqref{b2} and \eqref{b3} remain. The $(4;2)$ part of
\eqref{b2} gives
\bea
(\tilde{\dd}\mbox{Re}\Omega_{SU(3)}\mbox{Re}\Omega_{SU(2)ab}+2\mbox{Re}\Omega_{SU(3)}\mbox{Re}\Omega_{SU(2)c[a}Y^c_{\;\;\;b]})\w
e^a\w e^b=0.
\eea
Contracting the term in parentheses with $\mbox{Re}\Omega_{SU(2)ab}$,
we find that 
\bea
\tilde{\dd}\mbox{Re}\Omega_{SU(3)}&=&0,\\\label{b8}
\mbox{Re}\Omega_{SU(2)c[a}Y^c_{\;\;\;b]}&=&0.
\eea
Similarly for the $(4;2)$ part of \eqref{b3}. Now, equations \eqref{b7} and \eqref{b8} imply that $Y^{ab}$ must be
antisymmetric, and, with orientation
$\mbox{Vol}_{SU(2)}=\frac{1}{2}J_{SU(2)}\w J_{SU(2)}$, anti-selfdual
in the indices $a,b$. Therefore, regarded as the components of an
$SU(2)$ two-form, the $Y^{ab}$ lie in the adjoint, and this implies
that they may be set to zero locally by performing an $SU(2)$ rotation of the
$789\sharp$ directions, while preserving the metric, $J_{SU(2)}$ and
$\Omega_{SU(2)}$. Thus we have
\bea
\dd e^a=\hat{\dd}e^a.
\eea
Therefore we may always locally choose the frame on $\mathcal{M}_{SU(2)}$ to
be independent of the coordinates, and coordinate differentials, of
$\mathcal{M}_{SU(3)}$.    

Next consider the $(2;4)$ parts of \eqref{b2}, \eqref{b3}. From these
we find 
\bea
\Omega_{SU(2)}\w U^A=0.
\eea
Finally, from the (3;3) part of \eqref{b2} we get
\bea
3\mbox{Re}\Omega_{SU(3)D[AB}V^D_{\;\;\;C]}\mbox{Re}\Omega_{SU(2)}-\mbox{Re}\Omega_{SU(3)ABC}\hat{\dd}\mbox{Re}\Omega_{SU(2)}=0.
\eea
Contracting this equation with $\mbox{Re}\Omega_{SU(3)ABC}$, using
\eqref{b9}, we find that
\bea
\mbox{Re}\Omega_{SU(3)D[AB}V^D_{\;\;\;C]}&=&0,\\
\hat{\dd}\mbox{Re}\Omega_{SU(2)}&=&0.
\eea
Similarly for the $(3;3)$ part of \eqref{b3}. 

We have now exhausted all the torsion conditions, so let us summarise
what we have found. The conditions on the $SU(2)$ forms are
\bea
\dd J_{SU(2)}=\dd \Omega_{SU(2)}=0.
\eea
With $I_{SU(2)}$ defined as in the main text, the conditions on the
$SU(3)$ forms may be summarised as
\bea
\mbox{Vol}_{SU(2)}\w\dd(\D^{-1}J_{SU(3)}^2)=\dd \star J_{SU(3)}&=&0,\\
I_{SU(2)}\w U^A&=&0,\\
\Omega_{SU(3)D[AB}V^D_{\;\;\;C]}&=&0,\\
\tilde{\dd}\Omega_{SU(3)}&=&0.
\eea
It is readily verified that the last three equations may be combined
into
\bea
I_{SU(2)}\w\dd\Omega_{SU(3)}&=&0.
\eea
Thus we obtain the results quoted in the main text.

\section{The $AdS$ limit of membranes on three folds and twofolds: technical details}\label{FF}
In this appendix, we will give more of the technical details of the
derivation of the $AdS$ limit of the wrapped brane supersymmetry
conditions for threefolds. The derivation for twofolds is very
similar, and is omitted. Our starting point is the equations
\bea\label{c1}
\dd(e^0\w\Omega_{SU(3)}\w I_{SU(2)})&=&0,\\\label{c2} 
\dd\star J_{SU(3)}&=&0,\\\label{c3}
F&=&-\dd(e^0\w J_{SU(3)}).
\eea
By using the transitive action of $SU(3)$ in six dimensions, we may
choose the part of the $AdS$ radial direction lying in
$\mathcal{M}_{SU(3)}$ to lie along $e^6$; then we have $\hat{w}=e^5$.
It is straightforward to derive \eqref{416} and \eqref{419} for the
flux by demanding that the only non-vanishing components of \eqref{c3}
contain a factor proportional to the $AdS$ volume form. Next look at
\eqref{c2}. After some manipulation, and using \eqref{416}, this may
be shown to be equivalent to
\bea\label{c8}
\mbox{Vol}_{S^3}\w\dd[\l^{-1}(1-\l^{3/2}\r^2)\mbox{Vol}_{\mathcal{M}_{SU(2)}}]=0.
\eea
We will see that this condition will in fact be implied by one of the
others we will derive. It remains to look at \eqref{c1}. We choose a
basis for the self-dual $SU(2)$ forms on the overall transverse space
according to
\bea
K^a=\dd R\w\s^a+\frac{1}{4}\e^{abc}\s^b\w\s^c,
\eea
where the $\s^a$ are the $SU(2)$ invariant one-forms. Then the part of
\eqref{c1} containing the real part of $\Omega_{SU(3)}$ gives
\bea\label{c7}
&&\dd\left[J^3\w\hat{w}-\frac{1}{\l^{3/4}\r}J^2\w\hat{\r}\right]\w\s^a\w\dd
r\nn&&-\dd\left[\l^{-1}\sss J^2\right]\w\frac{1}{2}\e^{abc}\s^b\w
\s^c\w\dd r=0,
\eea
with a similar equation from the $\mbox{Im}\Omega_{SU(3)}$ part. These
equations do not at first sight obviously imply those given in the
main text. However, observe that \eqref{416} implies that
$(\sigma^a\w\s^b)\lrcorner\dd J^1=(\sigma^a\w\s^b)\lrcorner\dd
\hat{w}=0$. This in turn implies $(\sigma^a\w\s^b)\lrcorner\dd
J^a=0$. Then, wedging \eqref{c7} with $\s^a$, we find
\bea
\dd\left[\l^{-1}\sss J^2\right]\w\mbox{Vol}_{S^3}=0,
\eea
and hence that
\bea
\dd\left[\l^{-1}\sss J^2\right]=\s^a\w\left(\s^a\lrcorner
\dd\left[\l^{-1}\sss J^2\right]\right).
\eea
But then we may write \eqref{c7} schematically as
\bea
A\w\s^a-B^a\w\mbox{Vol}_{S^3}=0,
\eea
and since $(\s^a\w\s^b)\lrcorner A=0$, we have $A=0$, $B^a=0$, and
therefore that
\bea
\dd\left[\l^{-1}\sss J^2\right]&=&0,\nn
\dd\left[J^3\w\hat{w}-\frac{1}{\l^{3/4}\r}J^2\w\hat{\r}\right]&=&0.
\eea
The first of these equations implies \eqref{c8}, and so we obtain the
results quoted in the main text.

\end{document}